\documentclass[epj]{svjour}
\usepackage{latexsym}
\usepackage{graphics}
\usepackage[latin1]{inputenc}
\usepackage{rotating}
\begin{document}
\renewcommand{\textfraction}{0.00000000001}
\renewcommand{\floatpagefraction}{1.0}
\title{Photoproduction of {\boldmath{$\pi^0\pi^0$-}}
 and {\boldmath{$\pi^0\pi^+$}}-pairs off the proton
from threshold to the second resonance region}
\author{ 
  F.~Zehr\inst{1},
  B.~Krusche\inst{1},  
  P.~Aguar\inst{2},
  J.~Ahrens\inst{2},
  J.R.M.~Annand\inst{3},
  H.J.~Arends\inst{2},
  R.~Beck\inst{2,4},
  V.~Bekrenev\inst{5},
  B.~Boillat\inst{1},
  A.~Braghieri\inst{6},
  D.~Branford\inst{7},
  W.J.~Briscoe\inst{8},
  J.~Brudvik\inst{9},
  S.~Cherepnya\inst{10},
  R.F.B.~Codling\inst{3},
  E.J.~Downie\inst{3},
  P.~Drexler\inst{11},
  D.I.~Glazier\inst{7},
  L.V.~Fil'kov\inst{10},
  R.~Gregor\inst{11},
  E.~Heid\inst{2},
  D.~Hornidge\inst{12},
  I. Jaegle\inst{1},
  O.~Jahn\inst{2},
  V.L.~Kashevarov\inst{10,2},
  A.~Knezevic\inst{13},
  R.~Kondratiev\inst{14},
  M.~Korolija\inst{13},
  M.~Kotulla\inst{1},
  D.~Krambrich\inst{2},  
  A.~Kulbardis\inst{5},
  M.~Lang\inst{2,4},
  V.~Lisin\inst{14},
  K.~Livingston\inst{3},
  S.~Lugert\inst{11},
  I.J.D.~MacGregor\inst{3},
  D.M.~Manley\inst{15},
  Y. Maghrbi\inst{1},
  M.~Martinez\inst{2},
  J.C.~McGeorge\inst{3},
  D.~Mekterovic\inst{13},
  V.~Metag\inst{11},
  B.M.K.~Nefkens\inst{9},
  A.~Nikolaev\inst{2,4},
  M.~Ostrick\inst{2},
  P.~Pedroni\inst{6},
  F.~Pheron\inst{1},
  A.~Polonski\inst{14},
  S.~Prakhov\inst{9},
  J.W.~Price\inst{9},
  G.~Rosner\inst{3},
  M.~Rost\inst{2},
  T.~Rostomyan\inst{6},
  S.~Schumann\inst{2,4},
  D.~Sober\inst{16},
  A.~Starostin\inst{9},
  I.~Supek\inst{13},
  C.M.~Tarbert\inst{7},
  A.~Thomas\inst{2},
  M.~Unverzagt\inst{2,4},
  Th.~Walcher\inst{2},
  D.P.~Watts\inst{7}
\newline(The Crystal Ball at MAMI, TAPS, and A2 Collaborations)
\mail{B. Krusche, Klingelbergstrasse 82, CH-4056 Basel, Switzerland,
\email{Bernd.Krusche@unibas.ch}}
}
\institute{Department of Physics, University of Basel, Ch-4056 Basel, Switzerland
  \and Institut f\"ur Kernphysik, University of Mainz, D-55099 Mainz, Germany
  \and SUPA, School of Physics and Astronomy, University of Glasgow, G12 8QQ, United Kingdom
  \and Helmholtz-Institut f\"ur Strahlen- und Kernphysik, University Bonn, D-53115 Bonn, Germany
  \and Petersburg Nuclear Physics Institute, RU-188300 Gatchina, Russia
  \and INFN Sezione di Pavia, I-27100 Pavia, Pavia, Italy
  \and School of Physics, University of Edinburgh, Edinburgh EH9 3JZ, United Kingdom
  \and Center for Nuclear Studies, The George Washington University, Washington, DC 20052, USA
  \and University of California Los Angeles, Los Angeles, California 90095-1547, USA
  \and Lebedev Physical Institute, RU-119991 Moscow, Russia
  \and II. Physikalisches Institut, University of Giessen, D-35392 Giessen, Germany
  \and Mount Allison University, Sackville, New Brunswick E4L3B5, Canada
  \and Rudjer Boskovic Institute, HR-10000 Zagreb, Croatia
  \and Institute for Nuclear Research, RU-125047 Moscow, Russia
  \and Kent State University, Kent, Ohio 44242, USA
  \and The Catholic University of America, Washington, DC 20064, USA
}
\authorrunning{F. Zehr et al.}
\titlerunning{Photoproduction of pion pairs off the proton}

\abstract{Precise total cross-sections and invariant-mass distributions
have been measured for photoproduction of pion pairs off the proton producing
$p\pi^0\pi^0$ and $n\pi^+\pi^0$ final states from the threshold region up
to 800 MeV incident photon energy. Additionally, beam helicity asymmetries have
been measured in the second resonance region (550 MeV - 820 MeV).
The experiment was performed at the tagged photon beam of the Mainz MAMI accelerator 
with the Crystal Ball and TAPS detectors combined to give an almost 4$\pi$ solid-angle 
electromagnetic calorimeter. The results are much more precise than any previous 
measurements and confirm the chiral perturbation theory predictions for 
the threshold behavior of these reactions. In the second resonance region,
the invariant-mass distributions of meson-meson and meson-nucleon pairs 
are in reasonable agreement with model predictions, but none of the models
reproduce the asymmetries for the mixed-charge channel.
} 
\maketitle

\section{Introduction}
The photoproduction of pion pairs has attracted a lot of interest over recent
years in view of different properties of the strong interaction. 
In particular, it can contribute to sensitive tests of chiral perturbation 
theory, to the investigation of non-ground-state decays of nucleon resonances 
via reaction chains like  $R\rightarrow R'\pi\rightarrow N\pi\pi$ 
($R$, $R'$ nucleon resonances, $N=n,p$), and to the analysis of in-medium 
properties of hadrons like the $\sigma$-meson.
The first topic is related to the threshold behavior of double pion production 
off the nucleon, which has been predicted by chiral perturbation theory  
\cite{Bernard_94,Bernard_96}. The second is important for the extraction of 
nucleon resonance properties from the electromagnetic excitations of the proton 
and the neutron. Finally, the interpretation of modifications of the pion-pion 
invariant-mass distributions in nuclear matter requires input from the elementary 
reactions off the free nucleon. 

Most of the recent progress in the theoretical treatment of the strong 
interaction comes from two different lines of research: (1) the development
of the numerical methods of lattice gauge theory, approaching the
non-perturbative regime downwards from large quark masses 
(see e.g. \cite{Duerr_08,Bulava_10,Edwards_11}) and (2) the methods of chiral 
perturbation theory (ChPT) \cite{Weinberg_79,Gasser_84}, 
extrapolating upwards from small momenta and vanishing quark masses. 
The latter, originally based on the approximate Goldstone boson nature of
the pion and later extended to the nucleon sector \cite{Jenkins_91,Bernard_92},
had a large success in correctly describing the threshold behavior of
photon-induced $\pi^0$ production off the proton. Agreement with experiment
\cite{Beck_90,Fuchs_96,Schmidt_05} was achieved due to the contribution of pion loop 
diagrams, which were not considered in the derivation of the older low-energy 
theorems (LET). Subsequently, different observables in single pion
photoproduction have been measured and interpreted in the framework of chiral
perturbation theory. Examples are the unitary cusp in the s-wave $E_{0^+}$ multipole of
$\pi^0$-photoproduction at the $\pi^+$ production threshold \cite{Bernstein_97}
and the polarizability of the charged pion \cite{Ahrens_05a} studied via the 
$\gamma p\rightarrow \gamma\pi^+ n$ reaction. 
In the meantime, it was  discovered \cite{Bernard_94,Bernard_96}
that the chiral loop diagrams play an even more important role for double pion 
production, opening a new door for ChPT tests. An unexpected
prediction was that, very close to threshold, the double $\pi^0$ channel is 
so much enhanced by loop corrections that its cross section rises above
the charged and mixed-charged channels. This is completely counterintuitive.
At higher incident photon energies the reactions involving pairs with charged 
pions ($\pi^0\pi^+$, $\pi^-\pi^+$) have much larger cross sections, and in 
single pion production in the threshold region, the $p\pi^0$ final state is
strongly suppressed with respect to $n\pi^+$. However, in double pion 
production, although contributions of order one vanish for neutral pion pairs, 
the corrections of order $M_{\pi}$ are predicted to be much larger than for 
the pairs with charged pions \cite{Bernard_96}. This result is stable 
when all next-to-leading order terms of order $M_{\pi}^2$ are considered. 
At this order also resonance contributions come into play via the
$N^{\star}N\pi\pi$ s-wave vertex. The largest contribution by far involves
the $P_{11}$(1440) resonance. Taking all contributions together, the 
leading-order chiral loop diagrams contribute roughly 2/3 of the total 2$\pi^0$ yield, 
making this channel the ideal testing ground for these contributions, 
which in most other reactions account only for small corrections. 
The calculations \cite{Bernard_96} predict the following threshold behavior
for $\gamma p\rightarrow p\pi^0\pi^0$:
\begin{equation}
\label{eq:bernard}
\sigma_{\mbox{tot}}(E_{\gamma})=0.6\;\; \mbox{nb}\cdot
\left(\frac{E_{\gamma}-E_{\gamma}^{\mbox{thr}}}{10\;\; \mbox{MeV}}\right)^2,
\label{eq:chpt}
\end{equation}
where $E_{\gamma}$ is the incident photon energy in MeV and 
$E_{\gamma}^{\mbox{thr}}=$ 308.8 MeV is the production threshold.  
The largest uncertainty in this prediction comes from the s-wave coupling 
of the $P_{11}$(1440) resonance, which was taken from an analysis of the 
$\pi N\rightarrow\pi\pi N$ reaction \cite{Bernard_95}. 
Equation (\ref{eq:chpt}) uses the central value of this coupling. If instead
the upper limit is used, the coefficient in the equation rises from 
0.6 nb to 0.9 nb. 

The threshold behavior of double $\pi^0$ production was also investigated
in the framework of the Gomez Tejedor-Oset (Valencia) model \cite{Gomez_96}. 
Here, it was found \cite{Roca_02} that the low-energy cross section is
considerably enhanced when final-state interaction (FSI) contributions are 
taken into account in addition to the pure tree-level treatment
in \cite{Gomez_96}. In particular the re-scattering of $\pi^+\pi^-$-pairs
produced via the $\Delta$-Kroll-Ruderman diagram into $\pi^0\pi^0$ gives
a considerable contribution. Altogether, the FSI effects almost double the
threshold cross section for the $2\pi^0$ channel, which can be regarded as a
kind of remnant of the chiral loop effects discussed above. However, 
very close to threshold, their final result is still smaller than the values 
predicted by the chiral theory.
   
At higher incident photon energies, double pion production is of interest in 
view of the decay of nucleon resonances. It is a well-known problem that far
fewer nucleon resonances have been observed in experiments than 
are predicted by quark models. A possible explanation could be the experimental
bias against states that couple only weakly to $N\pi$, introduced into the data 
base by the dominance of elastic pion scattering experiments. 
Photon-induced reactions can remove this bias for the initial state, but then 
final states other than $N\pi$ must be investigated. 
As discussed in detail in \cite{Krusche_03} three alternative decay modes of
nucleon resonances can be studied via double pion production: the sequential 
decay via an intermediate excited nucleon state (for all types of pion pairs), 
the decay into the $N\rho$ channel (for $\pi^0\pi^+$ or $\pi^+\pi^-$) and 
the emission of the $\sigma$-meson (for $\pi^0\pi^0$ or $\pi^+\pi^-$). 
The reaction $\gamma p\rightarrow p\pi^0\pi^0$ has been recently investigated in
detail in view of such resonance contributions with the Bonn-Gatchina (BoGa)
coupled-channel model up to incident photon energies of 1.3 GeV
\cite{Sarantsev_08,Thoma_08}. 

Much more data are available at lower incident 
photon energies in the second resonance region comprised of the $P_{11}$(1440), 
$S_{11}$(1535), and $D_{13}$(1520) states. Total cross sections and 
invariant-mass distributions of the $\pi\pi$- and the $\pi N$-pairs have been
measured with the DAPHNE and TAPS detectors at the MAMI accelerator in Mainz 
\cite{Braghieri_95,Haerter_97,Zabrodin_97,Zabrodin_99,Wolf_00,Kleber_00,Langgaertner_01,Kotulla_04},
at GRAAL in Grenoble (also linearly polarized beam asymmetry) 
\cite{Assafiri_03,Ajaka_07}, and with the CLAS detector at JLab (electroproduction) 
\cite{Ripani_03}. Recently polarization observables were also measured at 
the MAMI accelerator \cite{Ahrens_03,Ahrens_05,Ahrens_07,Ahrens_11,Krambrich_09} and 
at the CLAS facility \cite{Strauch_05}. However, it is somewhat surprising that,
despite all these efforts, the interpretation of the data is still controversial
even at low excitation energies, where only a few well-known resonances 
contribute. All models agree that the production of charged pions involves 
larger background contributions from non-resonant terms like pion-pole diagrams
or terms of the $\Delta$-Kroll Rudermann type. The extreme cases are the 
$\pi^+\pi^-$ final state, which at moderate incident photon energies is 
dominated by such contributions, and the $\pi^0\pi^0$ final state with only 
small background contributions. The latter is thus better suited for the study 
of resonance contributions. 

However, the reaction models do not even agree for the dominant 
contributions to double $\pi^0$ production. In the model of the Valencia group 
\cite{Gomez_96,Nacher_01,Nacher_02} the most important contribution comes from the 
$D_{13}(1520)\rightarrow \Delta\pi^0\rightarrow p\pi^0\pi^0$ reaction chain.
In the model of Laget and co-workers \cite{Assafiri_03} a much more prominent 
role is played by the $P_{11}(1440)\rightarrow N\sigma$ decay. The recent
Bonn-Gatchina analysis finds a large contribution from the $D_{33}(1700)$ state, 
which is almost negligible in the other models. The effective Lagrangian 
model from Fix and Arenh\"ovel (Two-Pion-MAID \cite{Fix_05}) is also dominated by 
the $D_{13}$ resonance, but strongly underestimates the total cross section at 
incident photon energies below 700 MeV. The results for the 
helicity dependence of the cross sections obtained from the 
experiments on the Gerasimov-Drell-Hearn sum rule 
\cite{Ahrens_03,Ahrens_05,Ahrens_07} show a dominance of the $\sigma_{3/2}$ 
component.
This is in line with the excitation of $D_{13}$ or $D_{33}$ resonances and 
limits possible $P_{11}$ contributions. 
 
The situation is even more complicated for the mix-charged $n\pi^0\pi^+$ 
and $p\pi^0\pi^-$ final states. Early model calculations in the framework of the 
Valencia or the Laget model (see e.g. \cite{Gomez_96}) could not even reproduce 
the total cross section. First experimental data for the invariant-mass distributions 
of the pion pairs \cite{Zabrodin_99,Langgaertner_01} indicated an enhancement at 
large invariant masses, which was interpreted as a possible contribution from 
$\rho$-mesons, e.g. via the $D_{13}(1520)\rightarrow N\rho$ decay. Introduction 
of this process into the models improved substantially the agreement with experiment 
\cite{Nacher_01,Nacher_02,Fix_05}. 

The behavior and interpretation of the low-energy double pion production reactions
are also relevant for different aspects of the much discussed in-medium properties 
of hadrons. The above mentioned coupling of the $D_{13}(1520)$ resonance to the
$N\rho$ channel has been discussed in \cite{Langgaertner_01} as a possible 
explanation of the strong suppression of the second resonance bump in total 
photoabsorption by nuclei \cite{Bianchi_94}. For the free nucleon this structure is 
dominated by this $D_{13}$ resonance. If it couples strongly to 
$N\rho$ and the $\rho$ is broadened or shifted to lower mass in nuclear matter, 
then also the nucleon resonance will be broadened and the bump structure will be 
suppressed for nuclei. The study of the line-shape of this resonance via single 
$\pi^0$ photoproduction off nuclei could not establish such a nuclear broadening
\cite{Krusche_01}. However, the interpretation of the data is greatly complicated 
by final-state-interaction effects.

The other important in-medium effect discussed in the context of low-energy
double pion production is the behavior of the $\sigma$-meson in nuclear matter. 
The masses of the $J^{\pi} = 0^-$ pion and its chiral partner the 
$J^{\pi} = 0^+$ $\sigma$-meson are very different in vacuum, which is a
well-known manifestation of chiral symmetry breaking. The masses should become
degenerate in the chiral limit, so that one naively expects a density dependence
of the mass due to partial chiral restoration effects. Results \cite{Bernard_87}
obtained in the framework of the Nambu-Jona-Lasinio model indicated indeed 
a strong drop of the $\sigma$ mass as a function of nuclear density which is already 
significant at normal nuclear density $\rho_0$ at which the pion mass remains 
stable. The predicted effect for the production of meson pairs off nuclei is a 
downward shift of the invariant-mass distributions of scalar isoscalar meson pairs.
This prediction has been experimentally investigated with pion- and photon-induced
double pion production reactions 
\cite{Bonutti_96,Bonutti_99,Bonutti_00,Camerini_04,Grion_05,Starostin_00,Messchendorp_02,Bloch_07}.
The CHAOS collaboration \cite{Bonutti_96,Bonutti_99,Bonutti_00,Camerini_04,Grion_05}
reported such a downward shift for $\pi^+\pi^-$ pairs with respect to $\pi^+\pi^+$
pairs from pion induced reactions. The Crystal Ball collaboration at BNL
\cite{Starostin_00} observed a low-mass enhancement of strength for heavy nuclei in 
$\pi^-$-induced $\pi^0\pi^0$ production. In photon-induced reactions, a 
downward shift of the invariant-mass distributions of $\pi^0\pi^0$ pairs with 
respect to $\pi^0\pi^{\pm}$ pairs has been measured by the TAPS collaboration 
\cite{Messchendorp_02,Bloch_07}.
However, intricate final-state-interaction effects \cite{Bloch_07}
complicate the interpretation of the results. They require detailed studies in the
framework of models, which have to rely on precise input for the elementary cross
sections off the free nucleon, in particular at low incident photon energies. 

The present paper reports a simultaneous, precise measurement of the total 
cross section and the invariant mass distributions of the 
$\gamma p\rightarrow p\pi^0\pi^0$ and $\gamma p\rightarrow n\pi^0\pi^+$ reactions
from as close as possible to threshold up to the second resonance region.
In addition, the use of a circularly polarized photon beam allowed the measurement 
of the beam-helicity asymmetry as a function of the azimuthal angle between the 
reaction and production planes. The results for this asymmetry, which are very 
sensitive to details of the reaction models, have already been published in a 
preceding Letter \cite{Krambrich_09}.

The paper is organized as follows. A brief description of the experimental setup
is given in Sec.~\ref{sec:experiments}. The data analysis, including calibration
procedures, particle identification, Monte Carlo simulations of the detector response, 
and estimates of systematic uncertainties, is summarized in Sec.~\ref{sec:analysis}. 
In Sec.~\ref{sec:results} (Results) first the threshold region is discussed,  
particularly in view of chiral perturbation theory predictions, followed by a
comparison of the results from several reaction models to the observables at 
higher incident photon energies.

\section{Experimental setup}
\label{sec:experiments}

The experiment was performed at the Mainzer Mikrotron (MAMI B) accelerator 
\cite{Herminghaus_83,Walcher_90} using the Glasgow tagging spectrometer 
\cite{Anthony_91,Hall_96}
and an almost $4\pi$ covering electromagnetic calorimeter combining the TAPS
\cite{Novotny_91,Gabler_94} and Crystal Ball (CB) \cite{Starostin_01} detectors. 
Here we give only a short summary of the main parameters of the setup. Details
can be found in \cite{Schumann_10}, which used the same data set for the 
investigation of the $\gamma p\rightarrow p\pi^0\gamma '$ reaction. 

\begin{figure}[htb]
\resizebox{0.47\textwidth}{!}{%
  \includegraphics{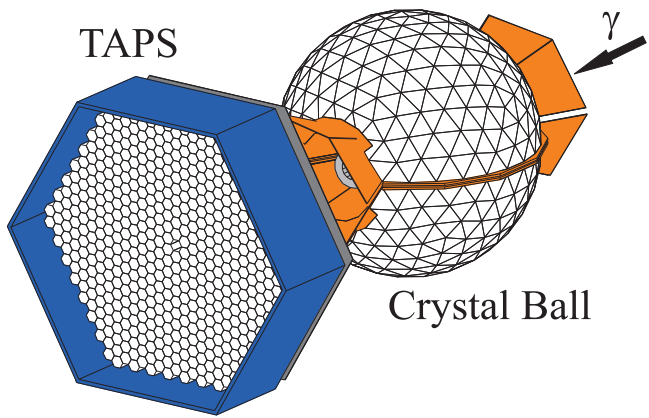}
}
\caption{Experimental setup of the electromagnetic calorimeter consisting of TAPS 
and Crystal Ball detector. Detectors for charged particle identification are mounted
inside the Crystal Ball around the target (PID and MWPCs) and in front of the TAPS
forward wall (TAPS Veto-detector).
}
\label{fig:setup}       
\end{figure}

A schematic drawing of the main components of the detector system is shown in 
Fig.~\ref{fig:setup}. The data were taken with a 4.8 cm long liquid hydrogen
target, which was mounted from the upstream side
in the center of the Crystal Ball. The Ball is composed of 672 NaI crystals
covering the full azimuthal range for polar angles between 
20$^{\circ}$ and 160$^{\circ}$. A {\bf P}article {\bf I}dentification {\bf D}etector 
(PID) \cite{Watts_04} (24 plastic scintillator bars of 31 cm length, 13 mm width
and 2 mm thickness) and two cylindrical multiple wire proportional chambers 
(MWPCs) \cite{Audit_91} were mounted inside the Ball in cylindrical geometry
around the target, covering the same polar angle range. The forward angular range 
from 20$^{\circ}$ down to 1$^{\circ}$ was covered by the TAPS detector 
\cite{Novotny_91,Gabler_94} with 510 BaF$_2$ crystals arranged as a hexagonal wall
and placed 1.75 m downstream from the target.
Each module of this detector had an individual plastic scintillator (5 mm thickness)
of the same hexagonal geometry in front for the identification of charged particles. 

The energy resolution for electromagnetic showers of both CB and TAPS is approximately
given by \cite{Starostin_01,Gabler_94}
\begin{equation}
\frac{\sigma_E}{E}\approx \frac{2 - 3\%}{\sqrt[4]{E/\mbox{GeV}}}\;\;\;.
\end{equation}
A more precise parameterization for TAPS can be found in \cite{Gabler_94}.
The angular resolution was better than 1$^{\circ}$ FWHM for TAPS and 
FWHM$_{\Theta}$ = 4.5$^{\circ}$ - 7$^{\circ}$ for polar and
FWHM$_{\phi}$ = FWHM$_{\Theta}$/sin($\Theta$) for azimuthal angles in the CB.  

The photon beam with energies up to 820 MeV was produced by bremsstrahlung from the
883 MeV electron beam of the MAMI accelerator. The energy of the incident 
photons was determined event-by-event by the Glasgow photon tagger 
\cite{Anthony_91,Hall_96} with an energy resolution of approximately 2 MeV full width. 
The electron beam was longitudinally polarized with a polarization degree of 
(82$\pm$5)\% so that the photon beam carried a circular polarization determined by 
the photon-energy dependent polarization transfer factor {\cite{Olsen_59}.
The incident photon flux was derived from the number of deflected electrons per 
tagger channel, which were counted with live-time gated scalers. The tagging 
efficiency, i.e. the fraction of correlated photons that pass through the collimator 
and impinge on the target, was periodically determined with special tagging efficiency 
runs. This was done by measuring directly the photon beam intensity with
a total absorbing lead-glass counter, which was moved into the photon beam at reduced 
intensity. During the normal data-taking runs the photon beam intensity was monitored 
in arbitrary units with an ionization chamber at the end of the beam line. After
normalization to the tagging efficiency runs these data were used to correct the
time dependence of the tagging efficiency, which however was quite stable (varying
between 30\% - 34\%).

\section{Data analysis}
\label{sec:analysis}

\subsection{Particle identification and reconstruction}

In general, electromagnetic showers produce signals in extended `clusters' of detector
modules in the calorimeters. The first step of the analysis therefore combined all
hits of adjacent crystals into `clusters' and determined their energy sums and their
energy weighted geometrical centers of gravity. In the next step these `clusters' 
were assigned to different particle types with the methods discussed below.   

\subsubsection{TAPS forward wall with veto-detector}

The response of the TAPS-detector to electromagnetic showers originating from
photons is discussed in detail in \cite{Gabler_94}. The separation between photons,
charged pions, and recoil nucleons in TAPS is based on three methods. The plastic
scintillators from the veto-detector distinguish between charged and neutral hits.
A hit was assigned as charged when the veto of any module in the cluster or the
veto of any neighbor of the central module (module with highest energy deposit) had
fired (even if the neighbor module itself had no signal above threshold). The latter
condition applies to cases where a charged particle with large impact angle passed
through the edge of a veto and then deposited its energy in a neighboring BaF$_2$ module.  

\begin{figure}[htb]
\resizebox{0.49\textwidth}{!}{%
  \includegraphics{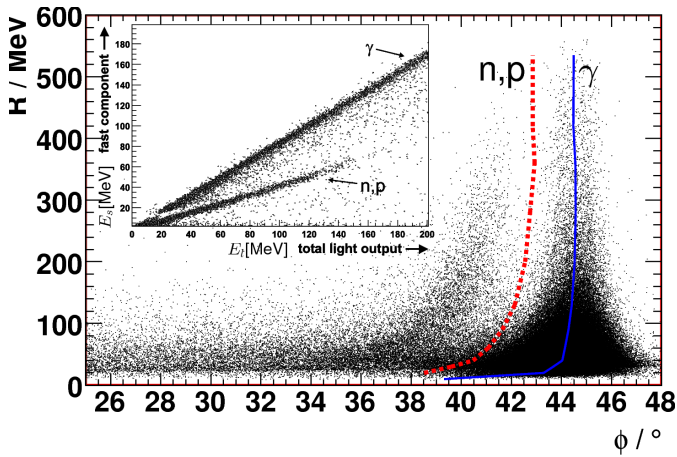}
  }
\caption{Pulse-shape analysis for the TAPS detector. The insert shows the signal integrated
over a short time gate versus the total signal. The main plot shows the 
pulse-shape signal in polar coordinates. Solid (blue) line: position of photon peak, 
dashed (red) line: 3$\sigma$ limit cut between photons and nucleons.
}
\label{fig:psa}       
\end{figure}

Photons and recoil nucleons can be separated by a pulse-shape analysis (PSA)
based on the scintillation properties of BaF$_2$. The crystals emit scintillation light
at two different wavelengths with very different decay constants ($\tau_f$ = 0.76 ns,
$\tau_s$ = 620 ns) and the intensity ratio of the two components depends on the nature 
of the incident radiation. This is routinely explored by integrating the output signals
over a short ($\approx$50 ns) and a long gate ($\approx$2 $\mu$s) period. The two 
energy signals were calibrated in a way that in a plot of long versus 
short-gate signal photons appear along the 45$^{\circ}$ line.
Since the fast component is suppressed for recoil nucleons, they appear at smaller angles
(see insert in Fig.~\ref{fig:psa}). For practical purposes the signals were 
parameterized in polar coordinates $R$, $\phi$ via
\begin{equation}
R  = \sqrt{(E^2_l+E^2_s)},\;\;\;\;
\phi  =  \tan^{-1}\left(\frac{E_{s}}{E_{l}}\right)\;\;,
\end{equation}
where $E_s$, $E_l$ are the short and long gate energy signals. The cut for the identification 
of photons was then defined in small slices of the spectra in the radius $R$ projected 
onto the $\phi$-axis. These spectra were fitted by Gaussian peaks plus a first order
background polynomial (see \cite{Zehr_10} for details). The solid (blue line) in 
Fig.~\ref{fig:psa} indicates the position of the Gaussian peaks and the dashed (red) line 
the 3$\sigma$ limit. Entries at the right hand side of the dashed line were accepted as 
photons.   

\begin{figure}[htb]
\resizebox{0.49\textwidth}{!}{%
  \includegraphics{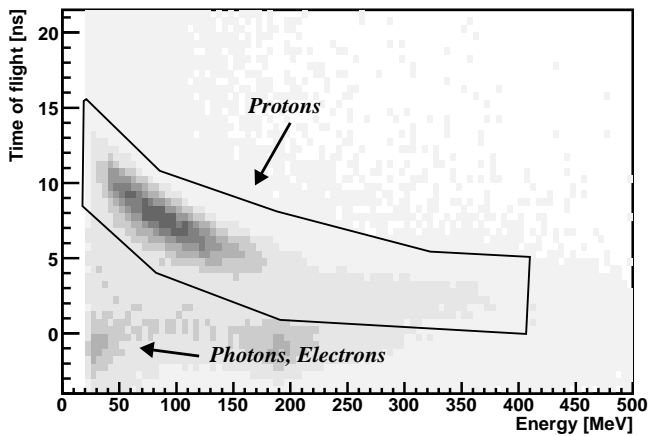}  
}
\caption{Time-of-flight versus energy analysis for nucleons detected in TAPS.
The offset of the time calibration is such that photons appear at zero time-of-flight.}
\label{fig:tofe}       
\end{figure}

A further possibility to separate protons and photons in TAPS is a time-of-flight 
versus energy analysis. This is demonstrated in Fig.~\ref{fig:tofe}. 
The selected events had exactly two photons in the Crystal Ball with an invariant mass 
close to the $\pi^0$ mass and one further hit in TAPS. This means that, apart from 
small background contributions, the reaction $\gamma p\rightarrow p\pi^0$ with the 
recoil proton in TAPS was selected.
In the figure the difference between the average time of the two photons and 
the time-of-flight for the hit in TAPS is plotted versus the energy deposited by the
hit in TAPS. Proton hits are confined in a well-defined curved zone. The indicated limits
define which hits are assigned as protons. The PSA and time-of-flight versus energy
analysis are complementary since the first is most efficient for large proton energies,
while the latter is optimal for the smallest energies. In principle, it is possible
to identify with these tools photons, neutrons, protons, and charged pions 
(see \cite{Bloch_07} for details, the pions form a separate band in the time-of-flight
versus energy spectra). However, as discussed in Sec. \ref{sec:reactions}, for the
present analysis they were only used to identify a very clean sample of photons 
in TAPS.

\subsubsection{Crystal Ball with PID and MWPCs}
\label{sec:MWPC}

Charged particles hitting the CB must traverse the PID and the MWPCs. The PID was used to
identify protons and charged pions. This was done by a $\Delta E - E$ analysis that
compared the differential energy loss of the particles in the PID to the total deposited 
energy in the CB. A typical spectrum is shown in Fig.~\ref{fig:pid}. The bands for 
protons and charged pions are clearly separated. 

\begin{figure}[htb]
\centerline{\resizebox{0.47\textwidth}{!}{%
  \includegraphics{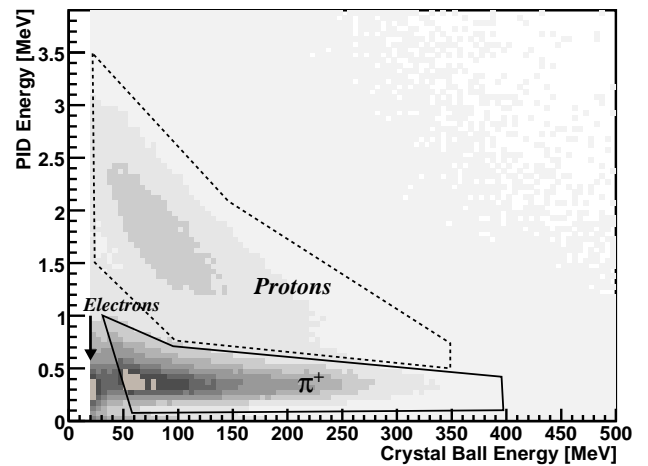}
}}
\caption{$\Delta E - E$-analysis using CB and PID. The energy deposition in the PID 
is plotted versus that in the CB.
}
\label{fig:pid}       
\end{figure}

The energy deposition of hadrons spreads over fewer detector modules than electromagnetic
showers and in many cases the energy deposit is confined to a single crystal. This means 
that the angular resolution for hadrons ($\approx10^{\circ}$ for $\Theta$) is worse than 
for photons for which the center of gravity of the extended cluster defines the impact 
point better than the detector granularity. Therefore, for charged particles the angular 
information from the CB was replaced by the tracking information delivered by the MWPCs 
whenever such information was available (if not the reconstruction from the CB clusters 
was used). Using the intersection points of the particle trajectory with the two MWPCs 
an angular resolution of FWHM$_{\Theta}$ = 3$^{\circ}$ - 5.5$^{\circ}$ and 
FWHM$_{\phi}$ = 3.3$^{\circ}$ was obtained. The angular resolution and the
detection efficiency of the MWPCs was experimentally determined with penetrating
cosmic muons and the reactions $\gamma p\rightarrow p\pi^0$ and 
$\gamma p\rightarrow n\pi^+$. For the latter two, events with protons, respectively
charged pions, identified by the $E - \Delta E$ analysis were selected and it was 
measured for which fraction of those events the MWPCs had responded. 
The total efficiency of the MWPCs determined this way was 90\% for protons 79\%
for $\pi^+$ mesons. Note, however, that this efficiency does not directly enter
into the total detection efficiency since $\pi^+$ mesons without MWPC information were
not discarded but had only a moderately lower angular resolution form the CB cluster
reconstruction (see section \ref{sec:xs} for details of the efficiency simulations).        

\subsection{Reaction identification}
\label{sec:reactions}

\subsubsection{The reaction $\gamma p\rightarrow p\pi^0\pi^0$}

For this reaction events with exactly four neutral hits and one or no candidate for the
recoil proton were accepted. Photons and proton (if detected) had to fulfill the above
discussed identification criteria. Detection of the recoil protons was not required for 
two reasons. Protons from reactions at low incident photon energies have low kinetic 
energies and were mostly stopped in the target or other material. Requiring proton
hits would have 
drastically reduced the overall detection efficiency and would have not allowed 
a measurement of the reaction close to threshold. At higher incident photon energies, 
protons were detected with good detection efficiencies. However, when detection of the 
proton is required, the simulation of the detection efficiency becomes more involved 
and more model dependent. This would have unnecessarily introduced an additional 
systematic uncertainty. 
   
\begin{figure}[htb]
\centerline{\resizebox{0.47\textwidth}{!}{%
  \includegraphics{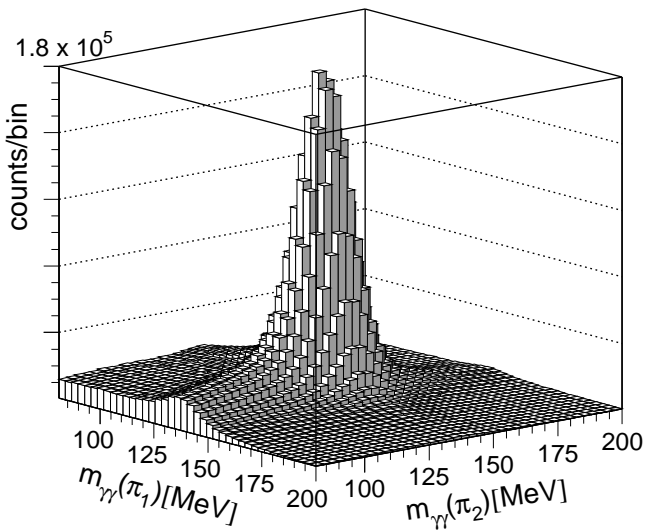}
}}
\caption{Invariant masses of the `best' combination of four photons to two
pairs for the $\gamma p\rightarrow p\pi^0\pi^0$ reaction.
}
\label{fig:invmass}       
\end{figure}

The accepted events were subjected to a combined invariant and missing mass analysis. 
In the first step, the four photons were combined into the three possible disjunct 
pairs. For each of the combinations the invariant mass of the two pairs was calculated
and the `best' combination was chosen with a $\chi^2$ test minimizing 
\begin{equation}
\chi^2 = \sum_{k=1}^{2}\frac{(m_{\gamma\gamma}(k)-m_{\pi^0})^2}{(\Delta
m_{\gamma\gamma}(k))^2}
\end{equation}
where $m_{\pi^0}$ is the  pion mass and the $m_{\gamma\gamma}$ are the invariant masses 
of the photon pairs with their uncertainties $\Delta m_{\gamma\gamma}$.
The two-dimensional spectrum of the invariant masses of the best combinations in 
Fig.~\ref{fig:invmass} shows a clear peak at the position of the $\pi^0$ invariant mass.
The small background under the peak was determined from a side-bin analysis and
subtracted. 

Subsequently, the nominal mass of the pion was used to improve the resolution.
Since the angular resolution of the detector system is much better than the energy 
resolution this was simply done by replacing the measured photon energies by
\begin{equation}
\label{eq:xform}
E'_{1,2}=E_{1,2}\frac{m_{\pi^0}}{m_{\gamma\gamma}}\;\;,
\end{equation}   
where $E_{1,2}$ are the  measured photon energies, $E'_{1,2}$ the re-calculated
energies, $m_{\pi^0}$ is the nominal $\pi^0$ mass, and $m_{\gamma\gamma}$ the
measured invariant mass. 
 
\begin{figure}[htb]
\centerline{\resizebox{0.47\textwidth}{!}{%
  \includegraphics{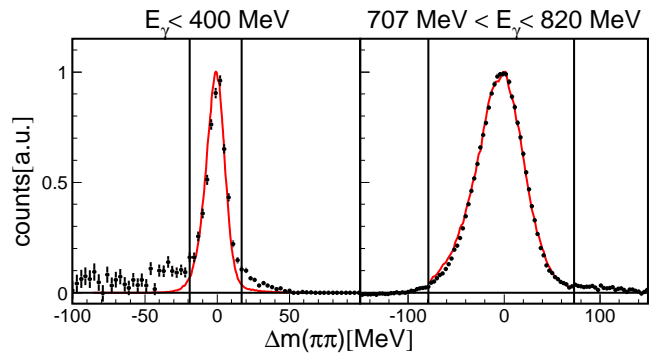}
}}
\caption{Missing mass spectra for the $\gamma p\rightarrow p\pi^0\pi^0$ reaction 
in the threshold range and at high incident photon energies. The solid (red) histograms
are GEANT simulations of the detector response. Vertical lines indicate the applied cuts.
}
\label{fig:2pi0_mm}       
\end{figure}

In the final step of the analysis the mass of the recoil proton, which was treated as
missing particle regardless of whether it was detected or not, was compared to 
the missing mass of the reaction using
\begin{equation}
\label{eq:2pimiss}
\Delta m(\pi\pi) = \left|P_{\gamma}+P_{p}- P_{\pi^0_1}
-P_{\pi^0_2}\right|
-m_p\ ,
\end{equation}
where $m_p$ is the nucleon mass, $P_{\gamma}$, $P_{p}$, 
$P_{\pi^0_{1,2}}$
are the four-momenta of the incident photon, the initial-state proton (which was at rest), 
and the produced $\pi^0$-mesons. The results for the two most critical energy regions -
close to threshold and at highest incident-photon energy - are shown in Fig.~\ref{fig:2pi0_mm}.
The critical point in the threshold region is the very low cross sections of double
$\pi^0$-production. However, the spectrum shows a rather clean peak. At higher incident 
photon energies, earlier experiments \cite{Haerter_97,Wolf_00}
were plagued by background from the $\eta\rightarrow 3\pi^0$ decay. The large solid angle
coverage of the present experiment results in a negligible probability to lose two out of
six photons so that this background did not contribute significantly. Only a
tiny remnant signal is visible around missing masses of +100 MeV.

In the final step, the kinetic energy and momentum of the recoil proton (no matter if 
detected or not) were reconstructed via the overdetermined reaction kinematics from 
the incident photon energy and the measured momenta of the two pions.

\subsubsection{The reaction $\gamma p\rightarrow n\pi^0\pi^+$}

For this reaction events with two or three neutral hits and a charged hit in the CB
were selected. Detection of the recoil neutron was allowed but not required. The charged 
hit in the CB had to pass the $\Delta E - E$ analysis as a charged pion. Charged pions 
in TAPS were not considered. The reason is that the identification
via the $\Delta E - E$ analysis using the PID was cleaner than the identification in 
TAPS where only time-of-flight versus energy could be used to distinguish charged pions 
from protons (charged pions would appear in Fig.~\ref{fig:tofe} between protons and 
photons; see \cite{Bloch_07} for details). Since protons misidentified as charged pions 
were the most important background source (see below), events with charged-pion 
candidates in TAPS were discarded. This introduced of course an additional systematic 
uncertainty into the simulation of the detection efficiency since a small part of the 
solid angle was not covered.   

In the first step of the analysis the invariant mass of the photon pair was calculated.
When three neutral hits had been detected in the CB, where neutrons and photons cannot be
distinguished, again the best combination was chosen. The resulting invariant mass
spectrum was very clean and a cut was applied for invariant masses between 115 MeV and 
160 MeV. As for the double $\pi^0$ channel the nominal mass of the $\pi^0$ was then used
to re-calculate the photon energies from Eq.~(\ref{eq:xform}).

Treating again the recoil nucleon as a missing particle, the missing mass can be
calculated from Eq.~(\ref{eq:2pimiss}), replacing one of the $\pi^0$-mesons by the 
$\pi^+$ and the final-state nucleon by the neutron. The result is shown in the center 
row of Fig.~\ref{fig:pi0piplus_mm}. The peaks around zero correspond to the 
$\gamma p\rightarrow n\pi^0\pi^+$ reaction. The background level is quite high, 
especially in the threshold region. The problem arises from protons from the 
$\gamma p\rightarrow p\pi^0$ reaction that leak in the $\Delta E - E$ analysis
(see Fig.~\ref{fig:pid}) into the $\pi^+$ region. The probability for this leakage is
small, however at energies below 400 MeV the cross section for single $\pi^0$
photoproduction is larger than for the $\gamma p\rightarrow n\pi^0\pi^+$ reaction by 
roughly three orders of magnitude. In order to reduce this background, first
the missing mass was calculated under the hypothesis of single $\pi^0$
photoproduction; i.e. the $\pi^+$ candidate was assumed to be a recoil proton
and the missing mass  
\begin{equation}
\label{eq:1pimiss}
\Delta m(\pi) = \left|P_{\gamma}+ P_{p}-P_{\pi^0}\right|
-m_p\ ,
\end{equation}
was formed. The result is shown in the upper row of Fig.~\ref{fig:pi0piplus_mm}.
In this spectrum, the background from single $\pi^0$ production sits in the peaks around
zero, while events from $\gamma p\rightarrow n\pi^0\pi^+$ appear at large missing masses.
For further analysis only the events in the shaded areas were accepted. The bottom row
of Fig.~\ref{fig:pi0piplus_mm} shows the two-pion missing mass after this cut on the
one-pion missing mass had been applied. These signals were practically background free 
with only a tiny contribution from $\eta\rightarrow \pi^0\pi^+\pi^-$ appearing at high 
incident photon energies and large missing mass. As for the double $\pi^0$ case the 
measured distributions agreed very well with the simulated line shapes.

\begin{figure}[htb]
\centerline{\resizebox{0.45\textwidth}{!}{%
  \includegraphics{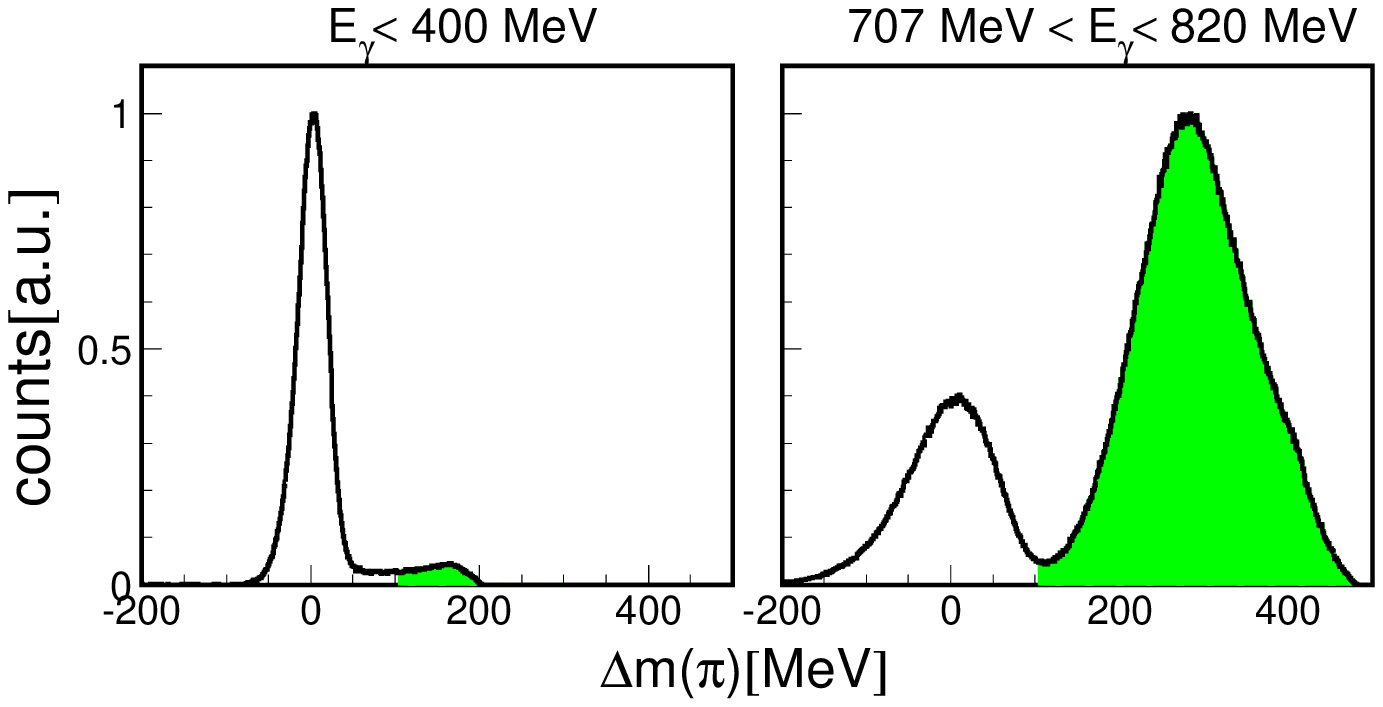}
}}
\centerline{\resizebox{0.45\textwidth}{!}{%
  \includegraphics{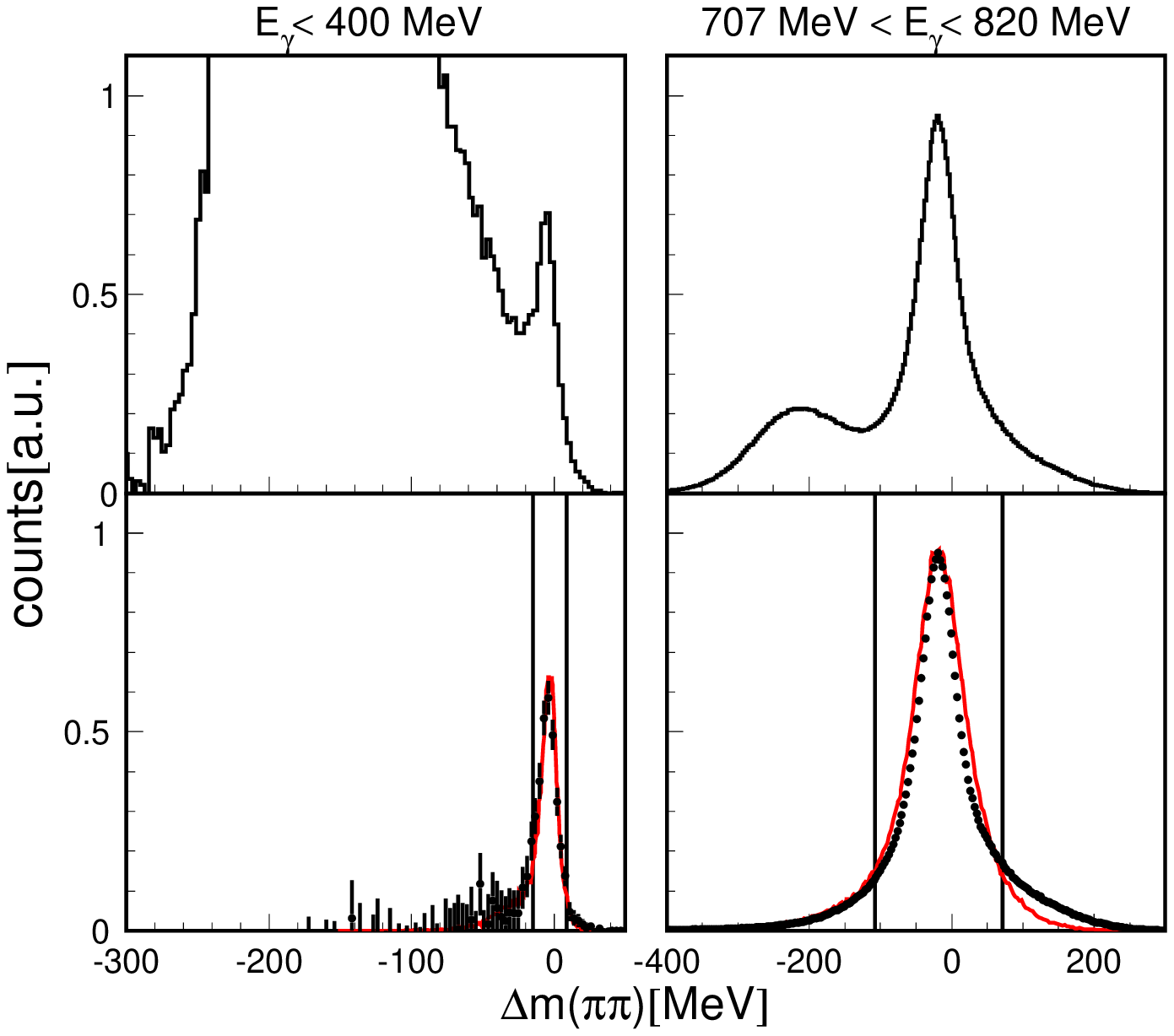}  
}}
\caption{Upper row: missing mass calculated from the $\pi^0$ kinematics for the 
hypothesis of the $\gamma p\rightarrow p\pi^0$ reaction. The peaks centered around 
zero are related to background from this reaction. The shaded areas were selected for 
further analysis of the $\gamma p\rightarrow n\pi^0\pi^+$ reaction. 
Center and bottom row: missing mass calculated from $\pi^0$ and $\pi^+$ kinematics 
for the hypothesis of the $\gamma p\rightarrow n\pi^0\pi^+$ reaction. Center: all events,
bottom: after cut on shaded areas in $\Delta m(\pi)$. Solid (red) histograms: simulation
of detector response. Vertical lines: applied cuts.
}
\label{fig:pi0piplus_mm}       
\end{figure}

At low incident photon energies there was another small, but significant background
component that could not be removed completely with the invariant-mass analysis. 
It arises
from the $\gamma p\rightarrow p\pi^+\pi^-$ reaction. Very slow $\pi^-$ mesons can be
stopped inside the liquid hydrogen target and then be captured by a proton to form 
pionic hydrogen. The pionic hydrogen can subsequently decay by charge exchange via the 
$\pi^- p\rightarrow n\pi^0$ reaction. Since this involves $\pi^-$ that started 
with very low momenta, their energy loss is not significant within the experimental 
resolution and they cannot be discriminated by the missing mass analysis. However, 
since the secondary $\pi^0$ mesons decay practically (within experimental resolution) 
at rest in the laboratory they can be eliminated in the pion kinetic energy spectra 
(peak at zero energy) or even in the opening angle spectra of the pion decay photons 
(peak at 180$^{\circ}$).

In the final step again the kinetic energy and momentum of the recoil neutron 
were reconstructed from the reaction kinematics, regardless whether the neutron was 
detected or not.

\subsection{Extraction of cross sections and systematic uncertainties}
\label{sec:xs}

The absolute normalization of the cross sections was based on the measurement 
of the incident photon flux, the target density, the two-photon-decay branching
ratio of the $\pi^0$ which is (98.823$\pm$0.034)\% \cite{PDG}, and the
efficiency of the detector system.

The photon flux was determined as explained in Section~\ref{sec:experiments}.
However, for the present experiment an unsolved problem occurred in the read-out 
electronics, which led to a staggered pattern in all types of 
$N_d(E_{\gamma})/N_{\gamma}(E_{\gamma})$ ratios, where $N_d$ are the counts for any 
reaction observed in the calorimeter and $N_{\gamma}$ is the photon flux. The pattern 
occurred in groups of four tagger channels with a maximum amplitude of $\pm$3.6\%. 
It was reduced to the average values by applying correction factors and a total 
systematic uncertainty of 5\% was adopted.

The target surface density was 0.201 nuclei/barn with a systematic uncertainty of 
2\%. Contributions from the target windows (in total 125 $\mu$m Kapton)
were determined with empty target measurements and subtracted. For the 
first part of the beam time, a build-up of ice on the downstream target window 
due to water permeation through the outer target tube was observed. The relative
thickness of this layer was monitored using reactions with protons at large polar
angles ($> 80^{\circ}$), which can only arise from the heavy 
nuclei in the target windows and the ice but not from reactions on the liquid 
hydrogen. The track
reconstruction with the MWPCs was used to measure the intensity of such events
from the different windows. Subsequently, the results from a measurement with 
a water target, normalized to the thickness of the ice layers, were used to 
subtract this background. For the second half of the beam time, this problem was 
eliminated by a modification of the target. The correction for the $\pi^0\pi^0$ 
cross section due to the ice layer for the first part of the beam time amounted 
to $\approx$ 10\%. We estimate the total systematic uncertainty due to this
correction is below the 2\% level. 

The systematic uncertainty due to the elimination of background via the invariant-
and missing-mass analysis (including the agreement between the observed line shapes 
and simulations) is estimated to be about the $\pm$3\% level for the $\pi^0\pi^0$ final
state and at $\pm$7\% for the $\pi^0\pi^+$ final state. It is larger for the latter
due to the additional background from stopped pions undergoing charge exchange and
because there is no invariant mass filter for the charged pion. 

The detection efficiency of the experimental setup was simulated with the 
GEANT3 program package \cite{Brun_86}. All details of the experimental setup 
(active detector components as well as support structures and other passive
materials) were included in the simulations and the results were tested in 
detail against known cross section data. The detector response to
electromagnetic showers induced by photons is very well known. The systematic 
uncertainty for single photon detection is so small, that even at very high 
statistical accuracy no significant deviations of cross section data constructed 
from the $\eta\rightarrow 3\pi^0\rightarrow 6\gamma$ decay from the world data 
base was observed \cite{McNicoll_10}. This was exploited to increase the statistical 
quality of $\eta$-production data by simultaneous measurements of the $2\gamma$- 
$6\gamma$-decay modes (see e.g. \cite{Pheron_12}), again without observation of
any systematic differences. For the charged pions, in addition the MWPCs must be 
considered, which are not routinely included in the GEANT simulation. They were 
treated in the following way. Efficiency and angular resolution of the chambers 
were experimentally determined (see sec. \ref{sec:MWPC}). In the simulation,
position information from the chambers was generated according to these experimental
parameters from the, in the simulation exactly known, tracks of the charged pions.
Whenever such information was available it replaced in the analysis, like for the 
measured data, the position information from the CB cluster. If not, the MWPCs
were ignored. For the detection efficiency of charged pions, this is only a second 
order effect. The efficiency is mainly determined by the GEANT3 simulation of PID 
and CB. The quality of GEANT3 simulations for charged pions in a calorimeter has
been for example studied in \cite{Bloch_07}, using the $\eta\rightarrow \pi^0\pi^+\pi^-$
decay. Also the agreement between the line-shapes of measured and simulated data
(cf. Fig. \ref{fig:pi0piplus_mm}) demonstrates the high quality of the simulations. 

The dominant uncertainty of the detection efficiency is related to the event generators
used for the simulation, which should reflect the `true' kinematic correlations 
between the two pions. This is discussed in detail below. For the 
$\gamma p\rightarrow p\pi^0\pi^0$ reaction, two different event generators were used. 
The first generated events for a phase-space distribution of the $p\pi^0\pi^0$ final 
state (ps). The second generator, called `sequential' (seq), simulated the reaction chain 
$\gamma p\rightarrow \Delta(1232)\pi^0\rightarrow p\pi^0\pi^0$ using a realistic 
mass distribution for the $\Delta(1232)$. This was motivated by previous results
\cite{Haerter_97,Wolf_00}, indicating a significant contribution from
decays of $N^{\star}$ resonances to the $\Delta(1232)$ intermediate state, which
gives rise to invariant-mass distributions different from phase-space behavior.

\begin{figure}[thb]
\centerline{\resizebox{0.46\textwidth}{!}{%
  \includegraphics{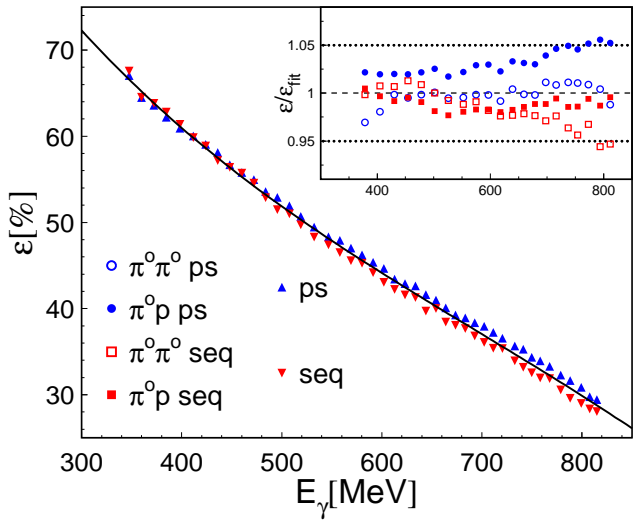}
}}
\caption{Double $\pi^0$ simulated detection efficiency for the 
$\gamma p\rightarrow p\pi^0\pi^0$ reaction. Main plot (blue) upward triangles:
average of efficiency extracted from pion-pion and pion-proton invariant mass
distributions for phase-space event generator, (red) downward triangles:
same for sequential event generator. (Black) solid line: average of (ps) and 
(seq). Insert: deviation of all four efficiency curves from the adopted average.  
}
\label{fig:effi1}       
\end{figure}
 
From both simulations the detection efficiency was extracted as 
$\epsilon(E_\gamma,m(\pi\pi)$) and as $\epsilon(E_\gamma ,m(\pi p)$),
i.e. as a function of incident photon energy $E_{\gamma}$ and the invariant mass
$m(\pi\pi)$ of the pion pairs and the invariant mass $m(\pi p)$ of the pion-proton
pairs. For the latter, the two identical pions were randomized.
Both types of invariant-mass distributions were then corrected with the 
appropriate detection efficiencies and the detection efficiency for the total cross 
section computed by integration over the distributions. The results are summarized in 
Fig.~\ref{fig:effi1}. Since the efficiencies obtained with the two different event 
generators differ only slightly, their average was used. As shown in the insert of 
Fig.~\ref{fig:effi1} the different results agree within $\pm$2\% - $\pm$3\% at 
lowest incident photon energies and within $\pm$6\% at highest incident photon 
energies. We therefore estimate a systematic uncertainty rising from 3\% at 310 MeV 
incident photon energy to 6\% at 800 MeV.
Some typical invariant-mass distributions are compared to the distributions used
in the event generators and to previous data in Fig.~\ref{fig:disc1}. 

In a more recent measurement with the CB/TAPS setup at MAMI-C, which will be published 
elsewhere \cite{Kashevarov_12}, data for $2\pi^0$ production off the proton was also 
taken up to higher incident photon energies (1.4 GeV) and analyzed in a different way,
using the kinematic-fit technique described in \cite{Prakhov_09,McNicoll_10}.
Due to this analysis and different trigger conditions, the detection efficiency
was quite different from the present experiment (see Fig.~\ref{fig:effi1}). 
For the range of incident photon energies discussed here, it was almost constant 
around 60\%. The MAMI-C measurements did not reach the same statistical precision in 
the threshold region as the present results, but they can serve as an independent 
cross-check for systematic uncertainties. 

The efficiency correction is more critical for the $\gamma p\rightarrow n\pi^0\pi^+$
reaction since in this case the charged pions were only accepted in the CB, which
excludes a small part of the reaction phase-space and must be extrapolated by the 
efficiency simulations. For this reaction three different event generators were 
used: phase-space and the reaction chains 
\begin{eqnarray}
\gamma p\rightarrow & \Delta^0(1232)\pi^+  &  \rightarrow n\pi^0\pi^+\\
\gamma p\rightarrow & \Delta^+(1232)\pi^0    &  \rightarrow n\pi^0\pi^+.
\end{eqnarray}

In a first step, the uncorrected invariant-mass distributions of the $n\pi^0$ 
and the $n\pi^+$ pairs were fitted with a superposition of the results from 
the three different simulations (the $\pi^0\pi^+$-pairs were much less sensitive 
to these reaction mechanisms). The relative contributions of these processes are
summarized in Fig.~\ref{fig:fractions}.

\begin{figure}[htb]
\centerline{\resizebox{0.46\textwidth}{!}{%
  \includegraphics{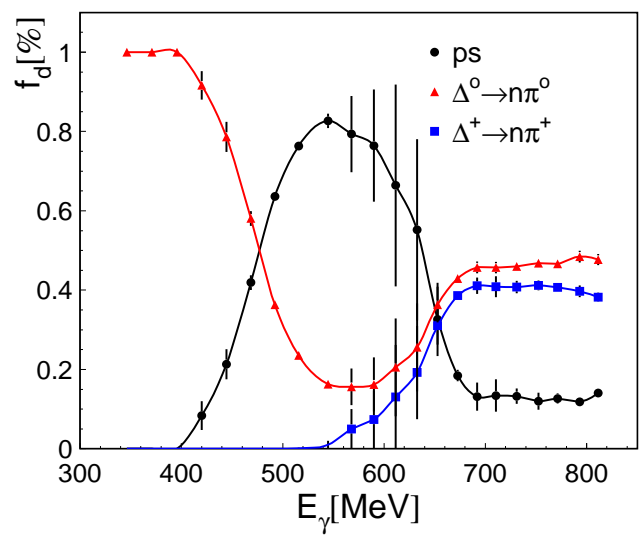}
}}
\caption{Relative contribution $f_d$ of phase-space (black dots),
$\Delta^0$-decays (red triangles) and $\Delta^+$-decays to the invariant-mass
distributions of $\gamma p\rightarrow n\pi^0\pi^+$. Symbols and curves represent 
the average of the fits for the $n\pi^0$ and $n\pi^+$ distributions; the error 
bars indicate their difference.
}
\label{fig:fractions}       
\end{figure}

It should be emphasized that this analysis was only intended to construct a 
realistic detection efficiency. It does not include minor contributions like
$\rho$-meson decays or interference terms. Nevertheless, the result
reflects properly the main features of the $\gamma p\rightarrow n\pi^0\pi^+$
reaction. Close to threshold it is dominated by the   
$\gamma p\rightarrow \Delta^0(1232)\pi^+\rightarrow n\pi^0\pi^+$ reaction chain.
This is due to contributions from pion-pole and $\Delta$-Kroll-Rudermann
background terms, where a $\Delta^0\pi^+$-pair is produced at the first vertex
and the $\Delta^0$ subsequently decays into $n\pi^0$. Such diagrams, 
with the two pions interchanged, do not contribute, since the incident photon 
couples only to charged pions. The dominance of 
$\gamma p\rightarrow \Delta^0(1232)\pi^+\rightarrow n\pi^0\pi^+$ at low energies 
is clearly visible in the invariant-mass distributions (see Fig.~\ref{fig:disc2}).
At the highest incident photon energies, contributions from sequential decays
of $N^{\star}$ resonances become important. Since, due to isospin invariance, the 
$N^{\star}\rightarrow \Delta^0\pi^+\rightarrow n\pi^0\pi^+$ and
$N^{\star}\rightarrow \Delta^+\pi^0\rightarrow n\pi^+\pi^0$ decays have the same
probability, the contributions from $\Delta^0$ and $\Delta^+$ decays become
comparable. 
\clearpage

\begin{figure*}[htb]
\centerline{\resizebox{1.0\textwidth}{!}{%
  \includegraphics{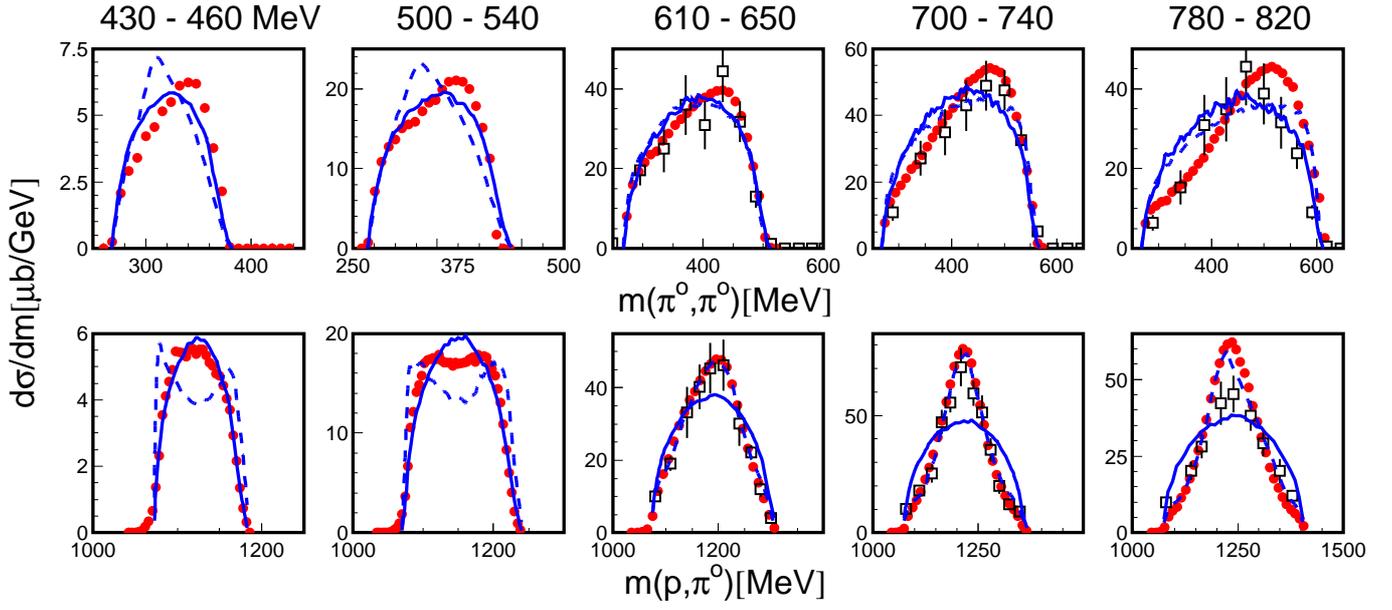}
}}
\caption{Typical invariant-mass distributions $m(\pi^0,\pi^0)$ and
$m(p, \pi^0)$ for different ranges of incident photon energy for 
$\gamma p\rightarrow p\pi^0\pi^0$. (Red) dots: present data,
(black) open squares: Wolf et al. \cite{Wolf_00}, 
solid lines: phase-space, dashed lines: sequential event generator.
}
\label{fig:disc1}       
\end{figure*}
\begin{figure*}[htb]
\resizebox{1.0\textwidth}{!}{%
  \includegraphics{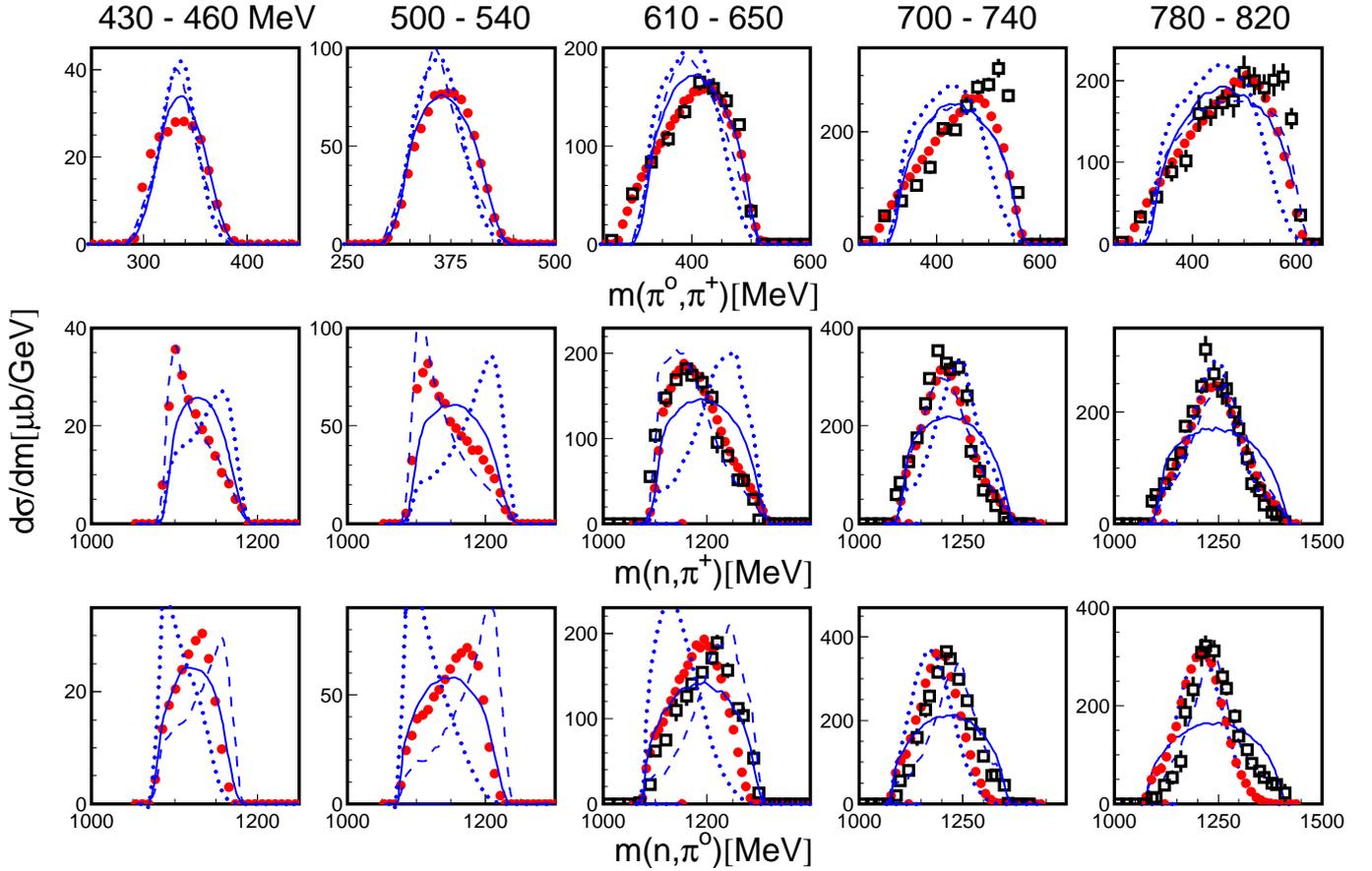}
}
\caption{Typical invariant-mass distributions (red) dots
$m(\pi^0,\pi^+)$, $m(n,\pi^+)$, $m(n,\pi^0)$ 
for $\gamma p\rightarrow n\pi^0\pi^+$.
(Black) squares: 
Langg\"artner et al. \cite{Langgaertner_01}. Solid lines: phase-space, 
dashed lines: $\gamma p\rightarrow \Delta^0(1232)\pi^+\rightarrow n\pi^0\pi^+$,
dotted lines: $\gamma p\rightarrow \Delta^+(1232)\pi^0\rightarrow n\pi^0\pi^+$
event generators.
}
\label{fig:disc2}       
\end{figure*}
\clearpage

In the intermediate energy range, where none of these processes
is dominant, phase-space behavior parameterizes phenomenologically the contributions
from many different diagrams. The results from the fits of the two different types
of invariant-mass distributions are in quite good agreement for most of the energy
range; only around $E_{\gamma}\approx 600$~MeV do larger deviations occur. 

\begin{figure}[htb]
\centerline{\resizebox{0.48\textwidth}{!}{%
  \includegraphics{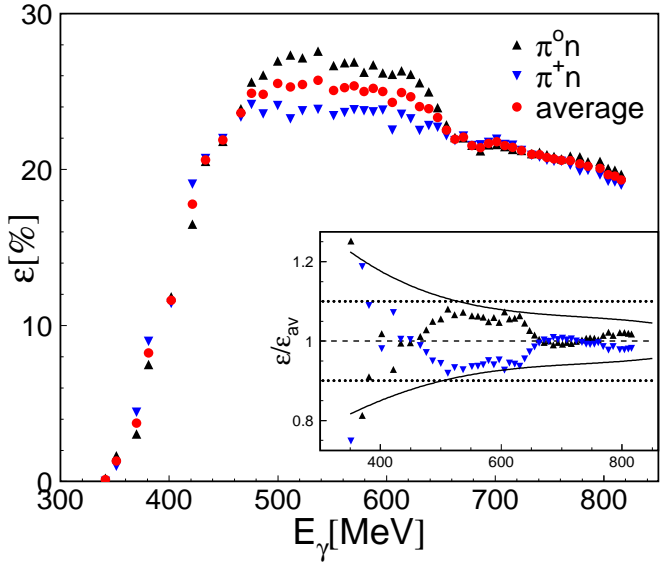}
}}
\caption{Detection efficiency for the $\gamma p\rightarrow n\pi^0\pi^+$ reaction.
(Black) upward triangles: from invariant-mass distributions of $\pi^0 n$ final 
state.
(Blue) downward triangles: from invariant-mass distributions of $\pi^+ n$ final 
state. (Red) dots: average. Insert: deviation from average. (Black) curves:
assumed systematic uncertainty of efficiency correction.
}
\label{fig:effi2}       
\end{figure}

For the final efficiency simulations, event generators with a corresponding mix 
of the three contributions were used and, as in the $\pi^0\pi^0$ case, the
detection efficiency was corrected as a function of the photon energy
$E_{\gamma}$ and the invariant mass of the particle pairs. As a check for 
systematic effects, this was done independently for the $\pi^0 n$ and $\pi^+ n$
pairs. The total detection efficiency was then again computed by integration.
The result is summarized in Fig.~\ref{fig:effi2}. The efficiencies obtained this 
way are in good agreement above $E_{\gamma}\approx 700$~MeV. They differ by up to 
$\pm$10\% at intermediate energies and by up to $\pm$20\% in the 
threshold region. For the final correction their average was used and the 
solid (black) curves in the insert of Fig.~\ref{fig:effi2} were assumed as
systematic uncertainties, i.e. $\pm$5\% at 800 MeV and $\pm$20\% at 350 MeV.

In contrast to double $\pi^0$ production, the detection efficiency becomes very small
in the threshold region because the charged pions are absorbed in the target or
other material. Therefore it was not possible to analyze this reaction close
to threshold.

\section{Results and discussion}
\label{sec:results}

\subsection{The threshold behavior}

The total cross section for double $\pi^0$ production in the threshold region 
is shown in Fig.~\ref{fig:2pi0_thres}. The comparison to previous results
\cite{Wolf_00,Kotulla_04} demonstrates the enormous progress achieved
in experimental precision over the last decade. While the first measurements of
this reaction \cite{Braghieri_95,Haerter_97} could not extract any meaningful 
results for incident photon energies below 400 MeV, the data by Wolf et al.
\cite{Wolf_00} from 2000 still have  statistical uncertainties on the 
100\% level, the results from Kotulla et al. \cite{Kotulla_04} reduced the 
uncertainties to the 50\% level, and the present results pushed them below 
the 10\% level (to the same magnitude as systematic uncertainties), allowing 
for the first time a stringent test of model predictions for this reaction.
The more recent MAMI-C data do not reach the same statistical precision, but do 
not show any systematic deviation from the present results. 

\begin{figure}[htb]
\resizebox{0.50\textwidth}{!}{%
  \includegraphics{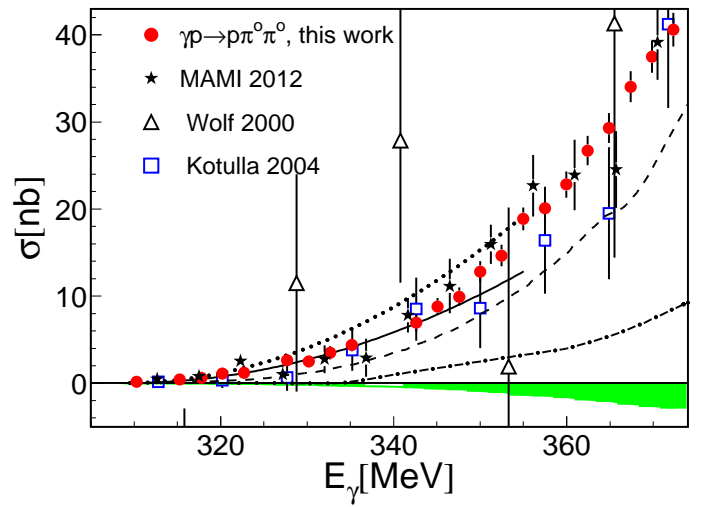}
}
\caption{Total cross section of the $\gamma p\rightarrow p\pi^0\pi^0$ reaction
in the immediate threshold region. (Red) dots: present measurement,
open (black) triangles \cite{Wolf_00}, open (blue) squares \cite{Kotulla_04},
(black stars) MAMI-C \cite{Kashevarov_12}. 
Solid line (dotted line): chiral perturbation theory prediction 
\cite{Bernard_96}, 
Eq.~(\ref{eq:bernard}) (dotted line: Eq.~(\ref{eq:bernard}) with 0.9 nb),
dashed line: prediction from Valencia model \cite{Roca_02},
dash-dotted: model by Fix and Arenh\"ovel \cite{Fix_05}. Shaded (green)
band at bottom: systematic uncertainty of present measurement. 
}
\label{fig:2pi0_thres}       
\end{figure}

The measured cross sections (Fig.~\ref{fig:2pi0_thres}) are in excellent agreement 
with the prediction from chiral perturbation theory \cite{Bernard_96}, using the central 
value for the s-wave coupling of the $P_{11}$(1440) resonance to the double pion 
channel (Eq.~(\ref{eq:bernard}) with 0.6 nb). The calculation in the framework 
of the Valencia model by Roca et al. \cite{Roca_02} somewhat underestimates the 
threshold data, whereas the results from the model of Fix and Arenh\"ovel 
\cite{Fix_05} (Two-Pion-MAID) are much lower.

\begin{figure}[tb]
\resizebox{0.48\textwidth}{!}{%
  \includegraphics{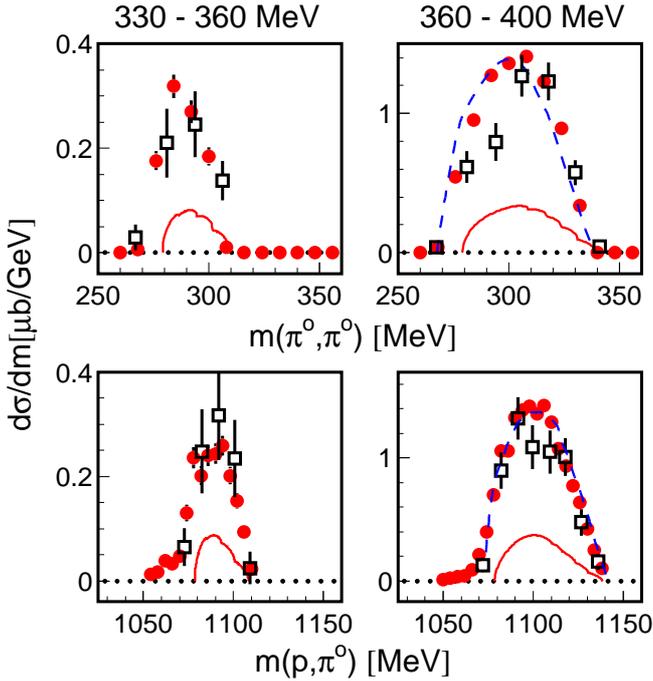}
}
\caption{Invariant-mass distributions for $\gamma p\rightarrow p\pi^0\pi^0$
in the threshold region. (Red) dots: present measurement, (black) open squares
Kotulla et al. \cite{Kotulla_04}. Dashed (blue) curves: phase-space, solid (red)
curves: model of Fix and Arenh\"ovel \cite{Fix_05}.
}
\label{fig:minv_thres}       
\end{figure}

\begin{figure}[h]
\centerline{
\resizebox{0.34\textwidth}{!}{%
  \includegraphics{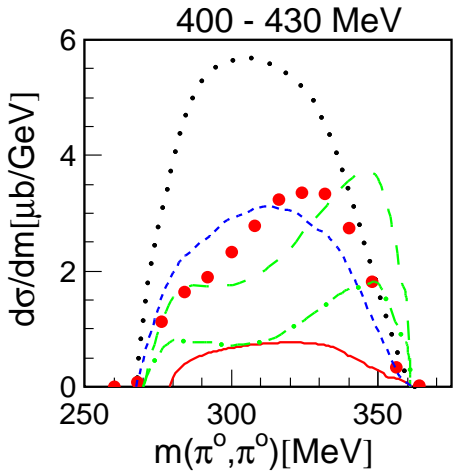}
}}
\caption{Invariant-mass distribution of the pion pairs from the 
$\gamma p\rightarrow p\pi^0\pi^0$ reaction. (Red) dots: present measurement.
Curves show model results from: Bonn-Gatchina model \cite{Sarantsev_08}
(black, dotted), Fix and Arenh\"ovel \cite{Fix_05} (red, solid), 
phase-space (renormalized in area, blue, dashed), Valencia model
\cite{Roca_02} (green, long-dashed with FSI, dash-dotted without FSI).
Note: calculation from \cite{Roca_02} is for $E_{\gamma}=430$~MeV.
It was re-normalized by the ratio of the total cross section at 430 MeV
and the average of the total cross section between 400 and 430 MeV.  
}
\label{fig:minv_roca}       
\end{figure}

\begin{figure}[hb]
\resizebox{0.50\textwidth}{!}{%
 \includegraphics{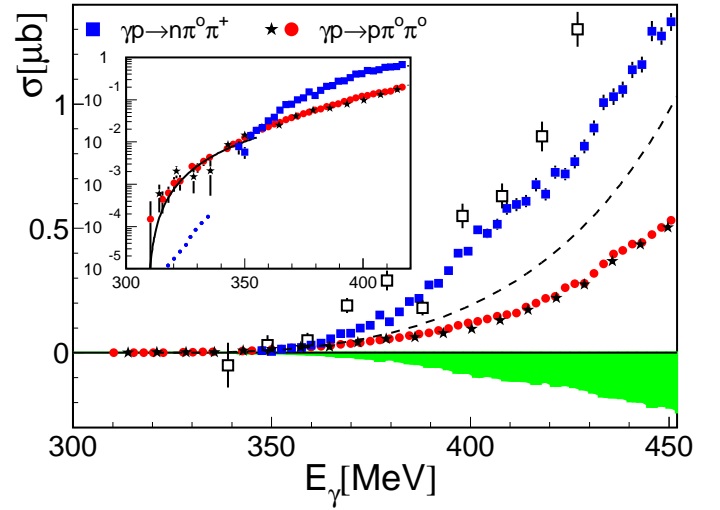}
}
\caption{Comparison of the threshold behavior of the  
$\gamma p\rightarrow p\pi^0\pi^0$ and $\gamma p\rightarrow n\pi^0\pi^+$ reactions.
(Red) dots: present measurement, (black) stars: MAMI-C for $\gamma p\rightarrow p\pi^0\pi^0$.
(Blue) squares present measurement, open (black) squares Langg\"artner et al. \cite{Langgaertner_01}
for $\gamma p\rightarrow n\pi^0\pi^+$.
Dashed curve: Two-Pion-MAID model \cite{Fix_05}
for $\gamma p\rightarrow n\pi^0\pi^+$. Shaded (green) band: systematic uncertainty
of the present $\gamma p\rightarrow n\pi^0\pi^+$ measurement.
Insert: comparison on logarithmic scale, solid (black) line: ChPT prediction for
$\pi^0\pi^0$ \cite{Bernard_96}, dotted (blue) line: ChPT prediction for $\pi^0\pi^+$
\cite{Bernard_94}.
}
\label{fig:pi0pic_thres}       
\end{figure}

Typical invariant-mass distributions for the threshold region are summarized
in Figs.~\ref{fig:minv_thres}, and \ref{fig:minv_roca}. Very close to threshold 
(see Fig.~\ref{fig:minv_thres}) pion-pion and pion-proton invariant masses
behave like phase-space (blue, dashed curves). At slightly higher incident
photon energies (see Fig.~\ref{fig:minv_roca}) the pion-pion distribution
is still similar to phase-space but develops some excess to large invariant 
masses. This is the typical energy range where in-medium modifications have 
been searched for in quasi-free production off heavy nuclei 
\cite{Messchendorp_02,Bloch_07}. The results from the reaction models are very
different for this energy range. The BoGa analysis \cite{Sarantsev_08}
resembles phase-space in shape but overestimates the data on an 
absolute scale. The prediction from the Two-Pion-MAID model \cite{Fix_05}
on the other hand underestimates the data but also has a different shape.
The calculation from Roca et al. \cite{Roca_02} is closest to the data and also
shows an accumulation of strength at high invariant masses, although this 
effect is slightly overestimated. 
The double-hump structure is due to an interference between the
isospin $I=0$ and $I=2$ amplitudes, which is large and destructive in the
Valencia model \cite{Roca_02}. It is also noteworthy that the pion-nucleon final-state
interaction in the $I=0$ channel has a large effect on the cross section 
in this model. It roughly doubles the result (compare dashed and dash-dotted
curves in Fig.~\ref{fig:minv_roca}) and effectively accounts for the
loop-corrections in chiral perturbation theory. This effect is not included in the
Two-Pion-MAID model and probably explains at least part of the missing strength
in this model.  

The threshold behaviors of $\gamma p\rightarrow p\pi^0\pi^0$ and 
$\gamma p\rightarrow n\pi^0\pi^+$ are compared in Fig.~\ref{fig:pi0pic_thres}. 
Unfortunately, it is not possible to measure $\gamma p\rightarrow n\pi^0\pi^+$
very close to threshold, since the low-energy charged 
pions do not reach the detector. The only alternative would have been to 
detect the neutral pion and the neutron and identify the reaction via 
overdetermined kinematics, but such events were not included in the 
trigger conditions (sum of energy deposited in Crystal Ball larger than 
60 MeV, combined multiplicity of hits in Crystal Ball and TAPS three or 
larger). Over the range that could be investigated the cross section is 
larger for the mixed-charge channel; however, at the lower limit of the 
accessible range there seems to be some indication for the cross-over 
predicted by chiral perturbation theory. A direct comparison to the 
predictions from chiral perturbation theory is not possible because they 
are limited to energies below 335 MeV (see insert of Fig.~\ref{fig:pi0pic_thres}). 
Also shown in Fig.~\ref{fig:pi0pic_thres} is the prediction from the 
Two-Pion-Maid model \cite{Fix_05}, which again underestimates the cross section, 
although not as dramatically as in the $\pi^0\pi^0$ case. 

\subsection{The resonance region}
\subsubsection{Total cross sections and invariant-mass distributions}

The total cross section for double $\pi^0$ production is shown in 
Fig.~\ref{fig:2pi0_tot} and the invariant-mass distributions are summarized in
Figs.~\ref{fig:dis1_2pi0} and \ref{fig:dis2_2pi0}. The agreement between the 
present data and the MAMI-C data \cite{Kashevarov_12} is excellent over the whole
energy range. 
\begin{figure}[htb]
\resizebox{0.50\textwidth}{!}{%
  \includegraphics{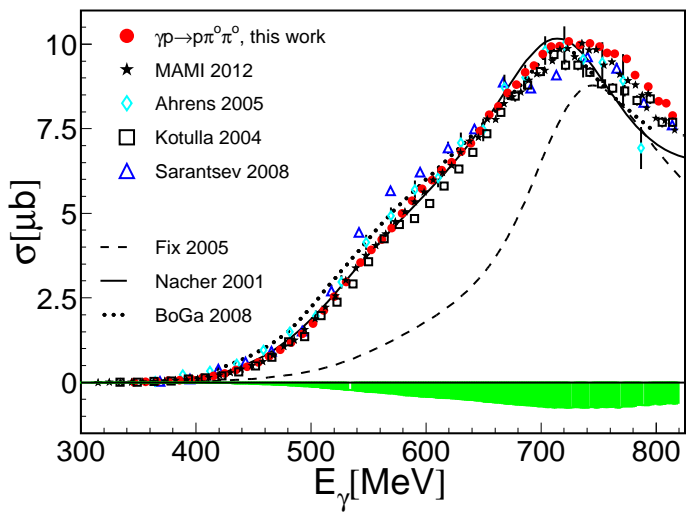}
}
\caption{Total cross section for the $\gamma p\rightarrow p\pi^0\pi^0$
reaction. (Red) dots: present measurement, (shaded (green) band: systematic uncertainty),
(black) stars: MAMI-C \cite{Kashevarov_12}, (black) squares: Kotulla et al.
\cite{Kotulla_04}, (blue) triangles: Sarantsev et al. \cite{Sarantsev_08},
(cyan) diamonds: Ahrens et al. \cite{Ahrens_05}. (Black) stars: new MAMI data.
Solid curve: Valencia model (Nacher et al. \cite{Nacher_01}),
dashed curve: Two-Pion-MAID (Fix and Arenh\"ovel \cite{Fix_05}), dotted curve: BoGa
(Sarantsev et al. \cite{Sarantsev_08}). 
}
\label{fig:2pi0_tot}       
\end{figure}

For the invariant mass distributions only the present results are shown,
since in most energy bins the MAMI-C data fall exactly on top of them.  
The total cross section
agrees within the systematic uncertainties with the earlier measurements with 
TAPS at MAMI \cite{Kotulla_04}, the GDH experiment at MAMI \cite{Ahrens_05}, 
and for most of the energy range (excluding a small region around 550 MeV) 
also with the Crystal Barrel experiment at ELSA \cite{Sarantsev_08}.
For energies above 700 MeV, the present results are slightly higher than
previous measurements, but also still within systematic uncertainties. 
Concerning the systematic effects it should be noted that only the present 
measurement and the MAMI-C experiment detected the two pions over the full 
reaction phase space with fairly large efficiency (30\% - 70\% for the present 
measurement (see Fig.~\ref{fig:effi1}), $\approx$60\% in the MAMI-C case) and 
thus did not need any model dependent extrapolations. This is reflected in the
excellent agreement of the detection efficiencies simulated with different event 
generators and applied in different ways (see Fig.~\ref{fig:effi1}).
Furthermore, since detection of the recoil proton was not required, only
the very well understood Monte Carlo simulation of electromagnetic showers
induced from photons was needed for the efficiency calculation.

The agreement with the Valencia model \cite{Nacher_01} and the BoGa analysis 
\cite{Sarantsev_08} is comparable and clearly better than with the 
Two-Pion-MAID model \cite{Fix_05}.
The BoGa coupled-channel model was fitted (in addition to many other channels)
also to the previous TAPS- and CBELSA double $\pi^0$ data \cite{Sarantsev_08}, 
so that reasonable agreement could be expected. Nevertheless, for incident 
photon energies below 600 MeV it overestimates the magnitude of the cross
section and does not agree well with the shape of the pion-pion invariant-mass
distributions (Fig.~\ref{fig:dis1_2pi0}). 
In the same energy range the
Two-Pion-MAID model \cite{Fix_05} largely underestimates the magnitude and strongly
disagrees with the shape of the pion-pion invariant-mass distributions.
The Valencia model without FSI \cite{Gomez_96,Nacher_01} is in quite good agreement 
with the data already at incident photon energies above 550 MeV 
(see Fig.~\ref{fig:dis1_2pi0}), while below 450 MeV the influence of FSI is large
(see Fig.~\ref{fig:minv_roca}). All three models reproduce
pion-pion and pion-proton invariant masses quite well at incident photon energies
above 700~MeV, i.e. in the range where the reaction is supposed to be dominated
by sequential resonance decays via the intermediate $\Delta$(1232), the signal of
which is clearly visible in the pion-proton invariant mass.  

The total cross section for $\gamma p\rightarrow n\pi^0\pi^+$ is shown in 
Fig.~\ref{fig:pi0pic_tot} and the invariant-mass distributions are summarized in
Figs.~\ref{fig:dis1_pi0pic} and \ref{fig:dis2_pi0pic}. Agreement between the present
and previous measurements is within their systematic uncertainties.
The difference in the peak maximum around 750~MeV (also within systematic 
uncertainties) is probably due to the different procedures used for the simulation 
of the detection efficiencies. The procedure used in the present work is described
in Sec.~\ref{sec:xs}, Langg\"artner et al. \cite{Langgaertner_01} used a simple
phase-space model, and Ahrens et al. \cite{Ahrens_03} used the Valencia model 
\cite{Gomez_96}. For this channel, agreement with
the results from the Valencia model \cite{Nacher_01} and Two-Pion-MAID 
\cite{Fix_05} is comparable and for Two-Pion-MAID much better than for the double
neutral channel. An analysis in the framework of the BoGa model is not yet
available. At low incident photon energies the pion-nucleon invariant-mass distributions
clearly show an enhancement of the $\Delta^0\rightarrow n\pi^0$ decay from
background terms (which is also reflected in an enhancement of small $n\pi^+$
invariant masses). 

\clearpage

\begin{figure*}[htb]
\resizebox{1.0\textwidth}{!}{%
  \includegraphics{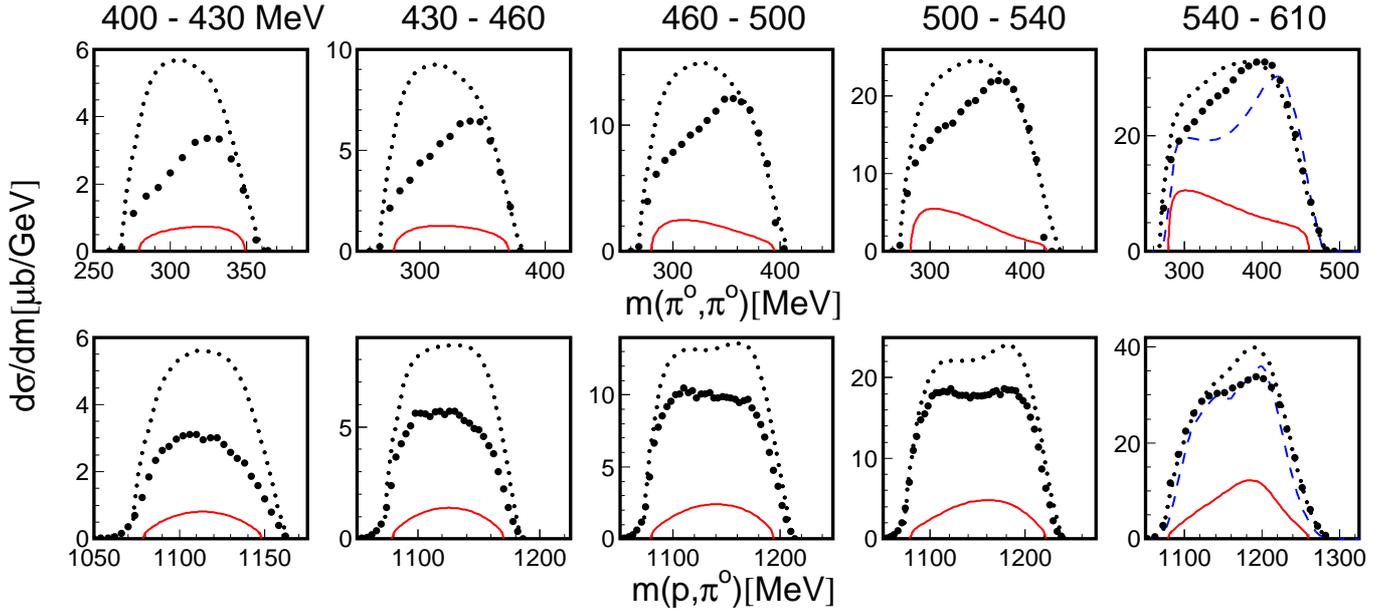}
}
\caption{Invariant-mass distributions of pion-pion and pion-proton pairs
for the $\gamma p\rightarrow p\pi^0\pi^0$ reaction for incident photon energies
from 400-610~MeV. (Black) dots: present measurement,
(red) solid curves: Two-Pion-Maid model by Fix and Arenh\"ovel \cite{Fix_05},
(black) dotted curves: BoGa analysis Sarantsev et al. \cite{Sarantsev_08},
(blue) dashed curves: Valencia model Nacher et al. \cite{Nacher_01} (only
available for the highest incident photon energy).
}
\label{fig:dis1_2pi0}       
\end{figure*}

\begin{figure*}[htb]
\resizebox{1.0\textwidth}{!}{%
  \includegraphics{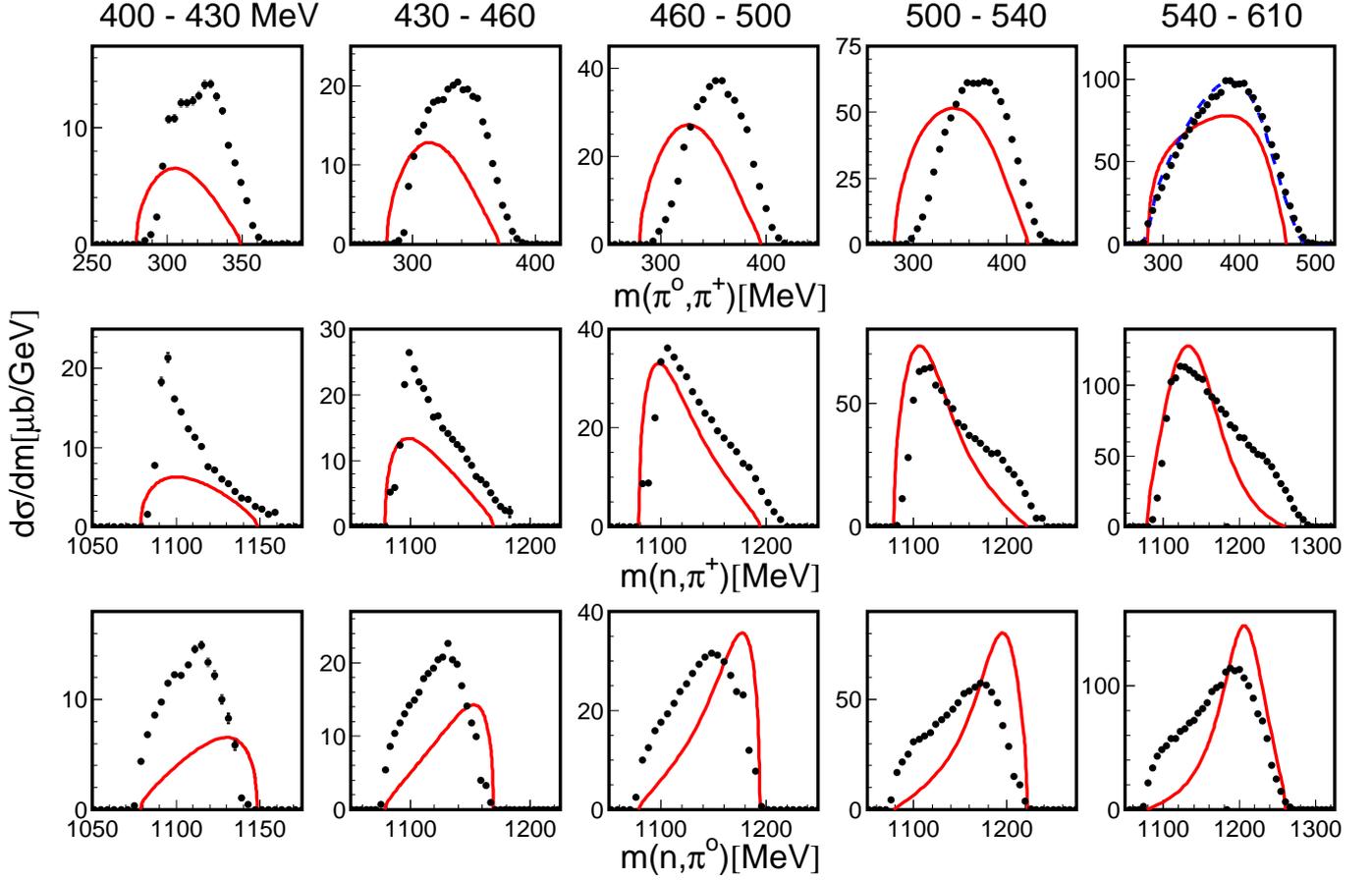}
}
\caption{Invariant-mass distributions of pion-pion and pion-neutron pairs
for the $\gamma p\rightarrow n\pi^0\pi^+$ reaction for incident photon energies
from 400-610~MeV. (Black) dots: present results,
(red) solid curves: Two-Pion-Maid model by Fix and Arenh\"ovel \cite{Fix_05},
(blue) dashed curve: Valencia model Nacher et al. \cite{Nacher_01} (only 
available the highest energy bin of pion-pion invariant mass).
}
\label{fig:dis1_pi0pic}       
\end{figure*}

\clearpage

\begin{figure*}[htb]
\resizebox{1.\textwidth}{!}{%
  \includegraphics{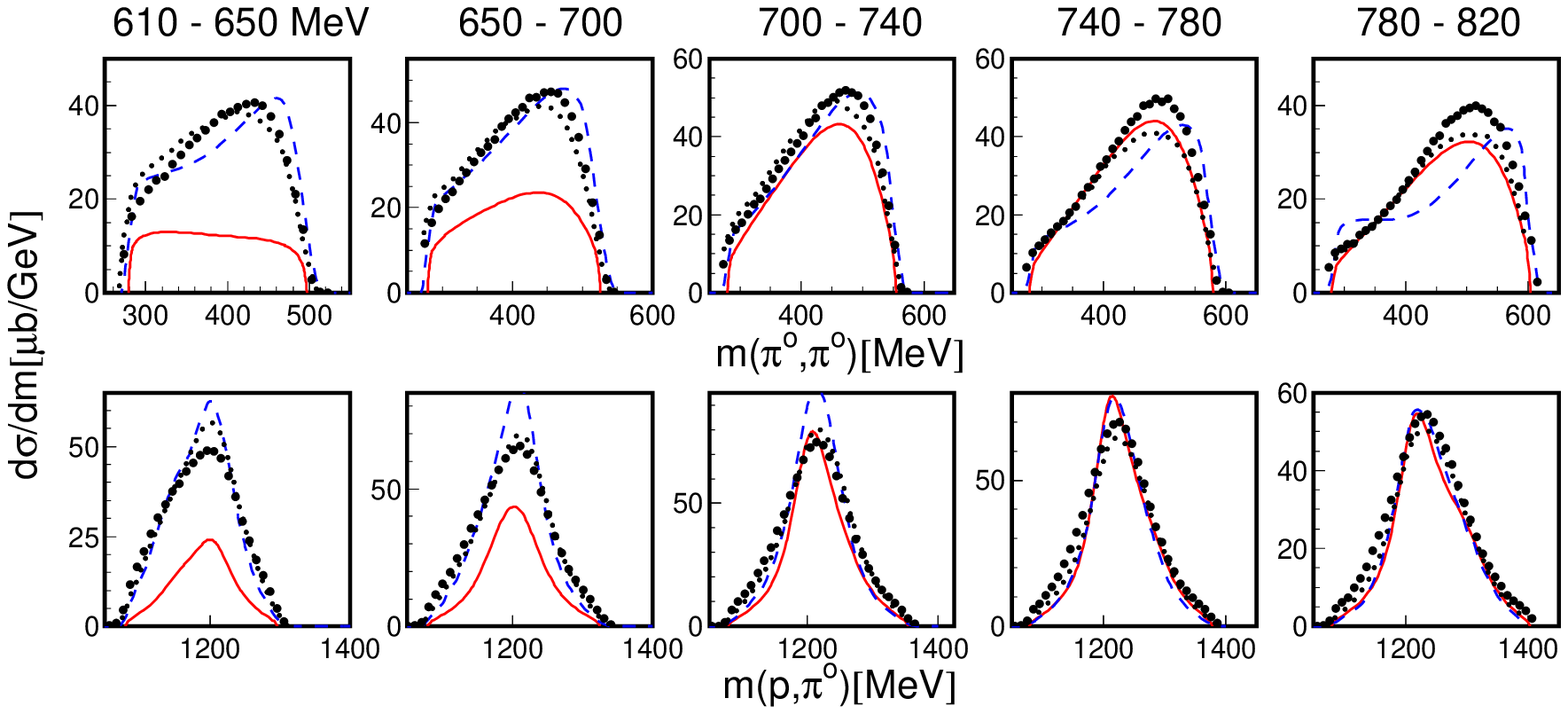}
}
\caption{Invariant-mass distributions of pion-pion and pion-proton pairs
for the $\gamma p\rightarrow p\pi^0\pi^0$ reaction for incident photon energies
from 610-820~MeV. (Black) dots: present measurement,
(red) solid curves: Two-Pion-Maid model by Fix and Arenh\"ovel \cite{Fix_05},
(black) dotted curves: BoGa analysis Sarantsev et al. \cite{Sarantsev_08},
(blue) dashed curves: Valencia model Nacher et al. \cite{Nacher_01}.
}
\label{fig:dis2_2pi0}       
\end{figure*}

\begin{figure*}[htb]
\resizebox{1.0\textwidth}{!}{%
  \includegraphics{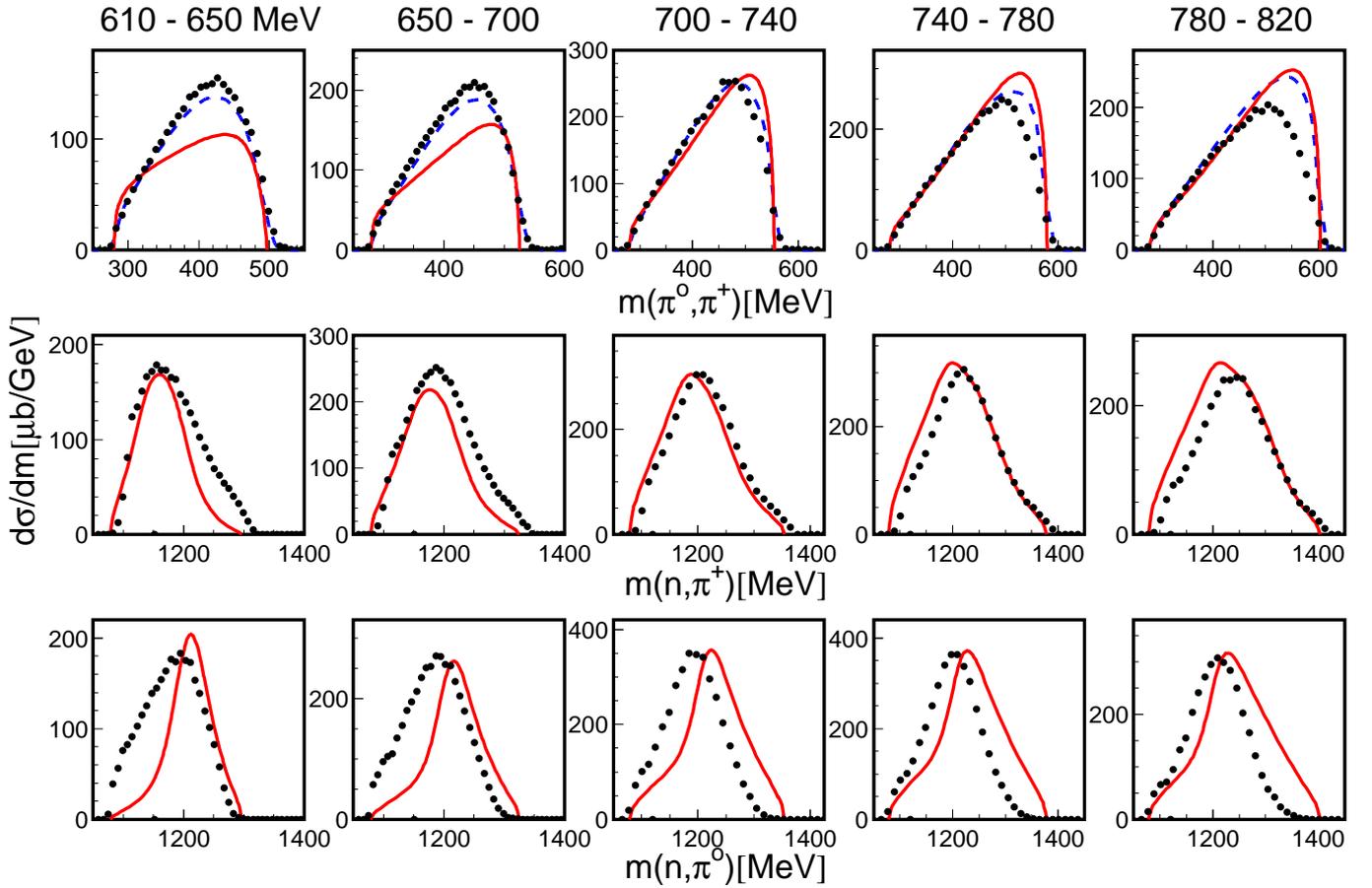}
}
\caption{Invariant-mass distributions of pion-pion and pion-neutron pairs
for the $\gamma p\rightarrow n\pi^0\pi^+$ reaction for incident photon energies
from 610-820~MeV. (Black) dots: present measurement,
(red) solid curves: Two-Pion-Maid model by Fix and Arenh\"ovel \cite{Fix_05},
(blue) dashed curves: Valencia model Nacher et al. \cite{Nacher_01} (only
available for pion-pion invariant mass).
}
\label{fig:dis2_pi0pic}       
\end{figure*}
\clearpage

At higher incident photon energies the $\Delta$ signal appears
in both pion-nucleon invariant masses as expected for sequential resonance decays.
The build-up of strength at large values of the pion-pion invariant mass at the
highest incident photon energies has been assigned to a contribution from
$\rho$-meson decays \cite{Zabrodin_99,Langgaertner_01,Nacher_01,Fix_05}.

\begin{figure}[htb]
\resizebox{0.50\textwidth}{!}{%
  \includegraphics{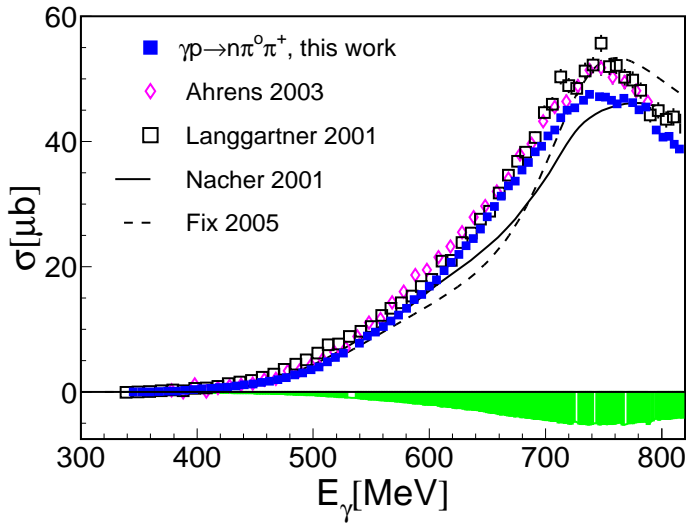}
}
\caption{Total cross section of the $\gamma p\rightarrow n\pi^0\pi^+$ reaction.
(Blue) squares: present measurement (shaded (green) band at bottom: systematic uncertainty), 
open (black) squares Langg\"artner et al. 
\cite{Langgaertner_01}, magenta diamonds: Ahrens et al. \cite{Ahrens_03}. 
Model results from Valencia (Nacher et al.
\cite{Nacher_01}) and Two-Pion-Maid (Fix and Arenh\"ovel \cite{Fix_05}).
}
\label{fig:pi0pic_tot}       
\end{figure}

\subsubsection{Beam helicity asymmetries}

Several model predictions \cite{Nacher_02,Fix_05,Roberts_05,Roca_05} indicate that
polarization observables are extremely sensitive to the reaction mechanisms. 
So far experimental results are scarce.
Beam asymmetries have been reported for the $\gamma p\rightarrow p\pi^0\pi^0$
\cite{Assafiri_03} and $\gamma n\rightarrow n\pi^0\pi^0$ \cite{Ajaka_07} reactions 
from the GRAAL experiment. The helicity dependence of the cross sections has been
measured with the Gerasimov-Drell-Hearn project at MAMI for
the $n\pi^+\pi^0$, $p\pi^0\pi^0$, $p\pi^+\pi^-$, and $p\pi^-\pi^0$ final states
\cite{Ahrens_03,Ahrens_05,Ahrens_07,Ahrens_11}.
The reaction models compare to all these data in a similar way as to the
unpolarized data. The predictions from the Valencia model \cite{Gomez_96,Nacher_01} 
and the Fix and Arenh\"ovel model \cite{Fix_05}, without agreeing in detail with the 
data, reproduce the main features of the split into $\sigma_{1/2}$ and $\sigma_{3/2}$ 
components. The strong dominance of the $\sigma_{3/2}$ component observed for the 
$p\pi^0\pi^0$ final state \cite{Ahrens_05} disfavors the Laget model \cite{Assafiri_03} 
where the dominant contribution is from the Roper resonance. In this situation it came 
as a surprise when the first measurement of the beam helicity asymmetry of the 
$\gamma p\rightarrow p\pi^+\pi^-$ reaction at the CLAS facility \cite{Strauch_05} 
produced results that could not at all be reproduced by any reaction model. 

\begin{figure}[htb]
\resizebox{0.50\textwidth}{!}{%
  \includegraphics{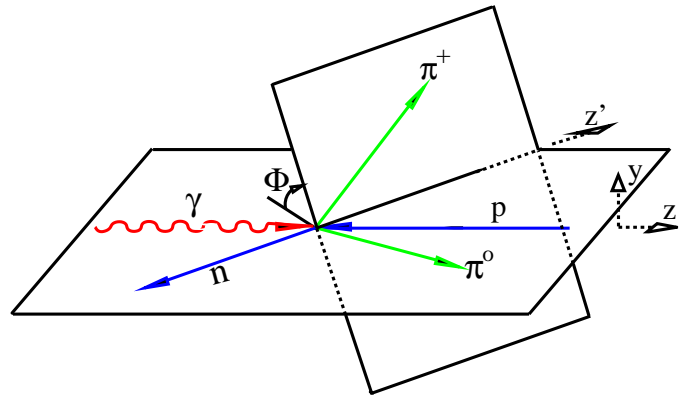}
}
\caption{Definition of the angle $\Phi$ between reaction plane (incoming photon and
outgoing nucleon) and production plane (pion pair). For identical pions
($\pi^0\pi^0$ pairs) the role of pion one and two has to be randomized. 
}
\label{fig:asym_def}       
\end{figure}

The present experiment has confirmed this result and for the first time measured 
this polarization observable for the $\pi^0\pi^0$ and $\pi^0\pi^+$ final states. 
Results have already been published in a preceding Letter \cite{Krambrich_09}.
Beam helicity asymmetries can be measured for reactions with at least three particles
in the final state with a circularly polarized photon beam on an unpolarized target.
The helicity asymmetry $I^{\odot}$ is defined by:
\begin{equation}
\label{eq:asym}
I^{\odot}(\Phi)=\frac{1}{P_{\gamma}}
	        \frac{d\sigma^{+}-d\sigma^{-}}{d\sigma^{+}+d\sigma^{-}}
	       =\frac{1}{P_{\gamma}}
                \frac{N^{+}-N^{-}}{N^{+}+N^{-}}\;\;,
\label{eq:circ}		
\end{equation}
where $d\sigma^{\pm}$ is the differential cross section for each of the two
photon helicity states, and $P_{\gamma}$ is the degree of circular polarization 
of the photons. For the extraction of the asymmetry 
$I^{\odot}(\Phi,\Theta_{\pi_1},\Theta_{\pi_2},...)$ at fixed kinematical parameters
the cross sections 
$d\sigma^{\pm}$ can be replaced by the raw count rates $N^{\pm}$ (right hand side of
Eq.~\ref{eq:asym}) since all normalization factors cancel in the ratio. In principal,
detection efficiency weighted count rates must be used for angle integrated asymmetries. 
However, due to the $\approx 4\pi$ coverage of the solid angle, detection efficiencies
were rather flat in phase space at fixed incident photon energy, so that the effect 
of the efficiency corrections on the asymmetries turned out to be negligible. 
The photon beam was produced from bremsstrahlung of longitudinally 
polarized electrons. In the energy range of interest, polarization degrees 
$P_{\gamma}$ between 60\% and 80\% were achieved.
The angle $\Phi$ between reaction and production plane (see Fig.~\ref{fig:asym_def}) 
was constructed in the same way as in the work of
Roca \cite{Roca_05}. For the $\pi^0\pi^+$ final state, the pions were ordered as shown
in the figure, i.e. $\Phi$ is the angle between the reaction plane and the part of
the production plane with the charged pion. For the double $\pi^0$ final state their 
assignment was randomized, which means that the asymmetry must obey 
$I^{\odot}(\Phi)=I^{\odot}(\Phi +\pi)$. This was taken into account in the modeling,
but not enforced in the data analysis. 

The advantage of this polarization observable is two-fold. It can be measured 
with good statistical quality (since only the electron beam must be polarized) and 
with small systematic uncertainties (since most uncertainties cancel in 
Eq.~(\ref{eq:asym})) and it is very sensitive to different contributions in the reaction 
models as has been demonstrated in \cite{Roca_02}. 

The results for the $\pi^0\pi^0$ and $\pi^0\pi^+$ channels are summarized in 
Figs.~\ref{fig:pi0pi0_asym} and \ref{fig:pi0pic_asym}. Parity conservation enforces that
$I^{\odot}(\Phi)=-I^{\odot}(2\pi-\Phi)$. This condition was not used in the analysis,
but as indicated in the figures it is almost perfectly respected by the measurements. 

\begin{figure}[htb]
\resizebox{0.50\textwidth}{!}{%
  \includegraphics{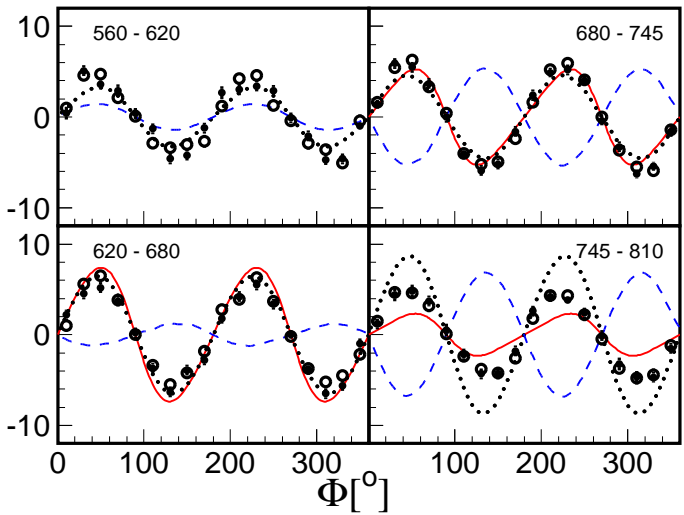}
}
\caption{Beam helicity asymmetry for $\gamma p\rightarrow p\pi^0\pi^0$ for four
different ranges of incident photon energy. Filled symbols: $I^{\odot}(\Phi)$, 
open symbols: $-I^{\odot}(2\pi-\Phi)$. (Red) solid curves: Two-Pion-MAID \cite{Fix_05},
(blue) dashed curves: Valencia model \cite{Roca_02}, (black) dotted curves: BoGa fit
\cite{Sarantsev_08}.
}
\label{fig:pi0pi0_asym}       
\end{figure}

\begin{figure}[htb]
\resizebox{0.50\textwidth}{!}{%
  \includegraphics{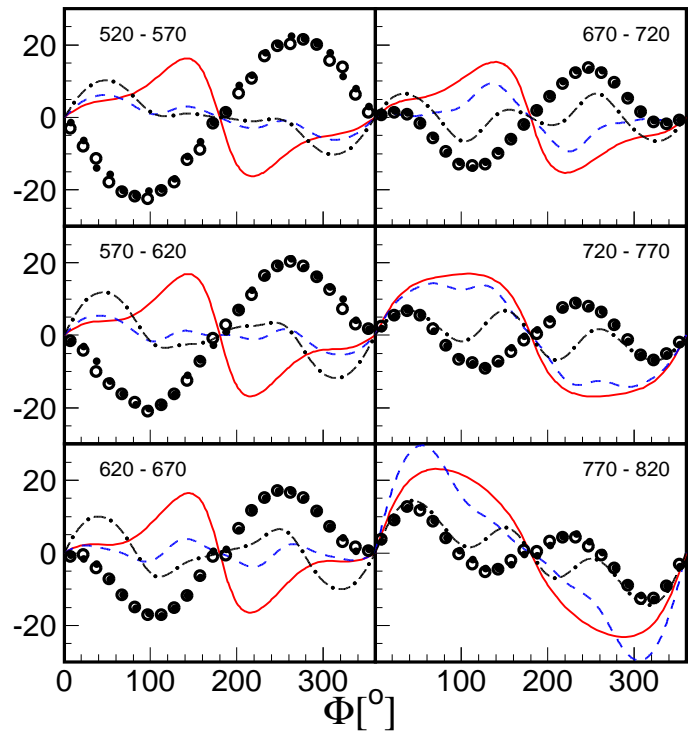}
}
\caption{Beam helicity asymmetry for $\gamma p\rightarrow n\pi^0\pi^+$ for six
different ranges of incident photon energy. Notation as in 
Fig.~\ref{fig:pi0pi0_asym} except (black) dash-dotted curves: Valencia model
\cite{Roca_02} without $\rho$-contributions. Results from BoGa are not available. 
}
\label{fig:pi0pic_asym}       
\end{figure}

The results for the double $\pi^0$ channel are in reasonable agreement with the
Two-Pion-MAID model \cite{Fix_05} and the BoGa analysis \cite{Sarantsev_08}.
The results from the Valencia model \cite{Roca_02} are completely out of phase 
with experiment.

In the case of the $\pi^0\pi^+$ final state, no model reproduces the experimental
results. This reaction is more complicated than $\gamma p\rightarrow p\pi^0\pi^0$,
which is dominated in the models by sequential resonance decays while 
$\gamma p\rightarrow n\pi^0\pi^+$ has additional contributions from the $\rho$
mesons and also stronger contributions from non-resonant background terms.
The predicted influence of such terms is shown in Fig.~\ref{fig:pi0pic_asym}
where for the Valencia model the full calculation and the result from a truncated 
model without the contributions from $\rho$-meson decays are compared.
Surprisingly, inclusion of the $\rho$ terms, which substantially improved the
agreement for the total cross section and the invariant-mass
distributions, has no positive effects for the asymmetry. The models do not even 
come close to the measurement. It is interesting to note, that for this reaction also the
relative contribution of $\sigma_{3/2}$ and $\sigma_{1/2}$ components \cite{Ahrens_03}
is quite different in the Valencia and the Fix and Arenh\"ovel model and for both of
them agreement with experiment is worse than for the other isospin channels. 
This means that the reaction mechanisms are still not understood in detail. 

\section{Summary and Conclusions}

Precise cross-sections and beam helicity asymmetries have been extracted for
the $\gamma p\rightarrow p\pi^0\pi^0$ and $\gamma p\rightarrow n\pi^0\pi^+$
reactions from the production thresholds up to the second resonance region. 

In the threshold region the results support strongly the predictions from chiral
perturbation theory \cite{Bernard_96}. The total cross section for double
$\pi^0$ production agrees with the ChPT prediction within statistical
uncertainties. It supports also the value used in the ChPT calculation for the
s-wave coupling of the $P_{11}$(1440) resonance to the double $\pi$ channel. 
The cross section calculated in the framework
of the Valencia model \cite{Roca_02}, taking into account $\pi N$ re-scattering
effects that resemble the loop corrections of ChPT, is also close to the measurement
but underestimates it slightly outside the experimental uncertainties. On the
other hand, the effective Lagrangian model (Two-Pion-MAID) from Fix and
Arenh\"ovel \cite{Fix_05} underestimates the threshold cross section typically by a
factor of five. This underlines the large importance of $\pi N$ re-scattering 
for the double $\pi^0$ channel. 

The results for $\pi^0\pi^+$ could not be directly compared to the chiral perturbation
prediction. There is a gap of $\approx$15 MeV between the upper energy limit of 
the theory prediction and the lower energy limit of the experimental 
sensitivity. Nevertheless, below incident photon energies of 400 MeV the $\pi^0\pi^+$ 
excitation function bends downwards, seems to cross the $\pi^0\pi^0$ data around 350 MeV,
and could approach the ChPT threshold prediction. But it is of course desirable to 
close the gap, which in terms of cross section spans almost two orders of magnitude,
either by more refined ChPT calculations reaching higher incident photon energies or 
by more sensitive experiments. Altogether, the experimental results clearly support 
the ChPT calculations four threshold production of pion pairs. 

The situation is much less clear at higher incident photon energies where the 
contributions from nucleon resonance decays become important. Fairly large
discrepancies between experiment and model results and between different models 
occur in the intermediate energy range from 400-650 MeV, where contributions 
from the $P_{11}$(1440) resonance have been discussed \cite{Sarantsev_08}. 
In the $\pi^0\pi^0$ channel the Two-Pion-MAID model \cite{Fix_05} strongly 
underestimates the magnitude of the cross section while the BoGa 
analysis \cite{Sarantsev_08} overestimates it. Both models disagree also 
in different ways with the shape of the invariant-mass distributions.
Agreement for Two-Pion-MAID with the $\pi^0\pi^+$ results is better as far as
the magnitude is concerned but also here the shape of the invariant-mass
distributions is not well reproduced. The results for total cross sections
and invariant-mass distributions from the Valencia model are at least in
reasonable agreement with experiment over the whole energy range. In the version
including $\pi N$ re-scattering, the $\pi^0\pi^0$ invariant-mass distributions
are quite well reproduced even at very low incident photon energies.

Finally, at the highest incident photon energies, in the second resonance 
region, all model analyses reproduce the absolute magnitude and the main
features of the shape of the invariant-mass distributions for both isospin 
channels. However, the models still do not agree on the relative importance 
of the different contributions, e.g. for the $D_{33}$(1700) resonance to
double $\pi^0$. The contribution of this state to $\pi^0\pi^0$ is very strong 
in the BoGa analysis, much smaller in Two-Pion-MAID, and almost 
negligible in the Valencia model. 

As a new tool, beam helicity asymmetries, which had been predicted to be very
sensitive to the details of the models, \cite{Roca_05} have been measured with 
high precision. The result is surprising. The Valencia model, which had the best
overall agreement with the measured cross section, could not reproduce this
observable for any of the isospin channels. Two-Pion-MAID and the BoGa analysis 
did much better for the double $\pi^0$ channel, but so far no model could
reproduce the results for the $\pi^0\pi^+$ channel (an analysis in the framework 
of BoGa is not available for this channel). Therefore we must conclude, that so 
far none of the available reaction models correctly reflects the details of 
double pion photoproduction in the resonance region. Further efforts on the theory
side are highly desirable.

\section{Acknowledgments}
We wish to acknowledge the outstanding support of the accelerator group 
and operators of MAMI. We thank A.~Fix, L.~Roca, A.V.~Sarantsev, and U.~Thoma
for useful discussions and the provision of the results from the different 
models. This work was supported by Schweizerischer Nationalfonds, Deutsche
Forschungsgemeinschaft (SFB 443, SFB/TR 16), DFG-RFBR (Grant No. 05-02-04014),
UK Science and Technology Facilities Council, STFC, European Community-Research 
Infrastructure Activity (FP6), the US DOE, US NSF and NSERC (Canada).
We thank the undergraduate students of Mount Allison University and The George Washington  
University for their assistance.


\begin{thebibliography}{99}
\bibitem{Duerr_08}        S. D\"urr et al.,                             Science                       {\bf 322},     1224    (2008).
\bibitem{Bulava_10}       J. Bulava et al.,                             Phys. Rev.                  D {\bf  82},   014507    (2010).
\bibitem{Edwards_11}      R.G. Edwards et al.,                          Phys. Rev.                  D {\bf  84},   074508    (2011).
\bibitem{Weinberg_79}     S. Weinberg                                   Physica                     A {\bf  96},      327    (1979).
\bibitem{Gasser_84}       J. Gasser, H. Leutwyler                       Ann. Phys.                    {\bf 158},      142    (1984).
\bibitem{Jenkins_91}      E. Jenkins, A.V. Manohar,                     Phys. Lett.                 B {\bf 255},      558    (1991).
\bibitem{Bernard_92}      V. Bernard, N. Kaiser, U.G. Meissner,         Nucl. Phys.                 B {\bf 383},      442    (1992).
\bibitem{Beck_90}         R. Beck et al.,                               Phys. Rev. Lett.              {\bf  65},     1841    (1990).
\bibitem{Fuchs_96}        M. Fuchs et al.,                              Phys. Lett.                 B {\bf 368},       20    (1996). 
\bibitem{Schmidt_05}      A. Schmidt et al.,                            Phys. Rev. Lett.              {\bf  87},   232501    (2005).
\bibitem{Bernstein_97}    A.M. Bernstein et al.,                        Phys. Rev.                  C {\bf  55},     1509    (1997).
\bibitem{Ahrens_05a}      J. Ahrens et al.,                             Eur. Phys. J.               A {\bf  23},      113    (2005).
\bibitem{Bernard_94}      V. Bernard et al.,                            Nucl. Phys.                 A {\bf 580},      475    (1994).
\bibitem{Bernard_96}      V. Bernard, N. Kaiser, U.G. Meissner,         Phys. Lett.                 B {\bf 382},       19    (1996).
\bibitem{Bernard_95}      V. Bernard, N. Kaiser, U.G. Meissner,         Nucl. Phys.                 B {\bf 457},      147    (1995).   
\bibitem{Gomez_96}        J.A. Gomez Tejedor and E. Oset,               Nucl. Phys.                 A {\bf 600},      413    (1996).
\bibitem{Roca_02}         L. Roca, E. Oset, M.J. Vicente Vacas,         Phys. Lett.                 B {\bf 541},       77    (2002).
\bibitem{Krusche_03}      B. Krusche and S. Schadmand,                  Prog. Part. Nucl. Phys.       {\bf  51},      399    (2003).
\bibitem{Sarantsev_08}    A.V. Sarantsev et al.,                        Phys. Lett.                 B {\bf 659},       94    (2008).
\bibitem{Thoma_08}        U. Thoma et al.,                              Phys. Lett.                 B {\bf 659},       87    (2008).
\bibitem{Braghieri_95}    A. Braghieri et al.,                          Phys. Lett.                 B {\bf 363},       46    (1995).
\bibitem{Haerter_97}      F. H\"arter et al.,                           Phys. Lett.                 B {\bf 401},      229    (1997).
\bibitem{Zabrodin_97}     A. Zabrodin et al.,                           Phys. Rev.                  C {\bf  55},    R1617    (1997).
\bibitem{Zabrodin_99}     A. Zabrodin et al.,                           Phys. Rev.                  C {\bf  60},   055201    (1999).
\bibitem{Wolf_00}         M. Wolf et al.,                               Eur. Phys. J.               A {\bf   9},        5    (2000).
\bibitem{Kleber_00}       V. Kleber et al.,                             Eur. Phys. J.               A {\bf   9},        1    (2000).
\bibitem{Langgaertner_01} W. Langg\"artner et al.,                      Phys. Rev. Lett.              {\bf  87},   052001    (2001).
\bibitem{Kotulla_04}      M. Kotulla et al.,                            Phys. Lett.                 B {\bf 578},       63    (2004).
\bibitem{Assafiri_03}     Y. Assafiri et al.,                           Phys. Rev. Lett.              {\bf  90},   222001    (2003).
\bibitem{Ajaka_07}        J. Ajaka et al.,                              Phys. Lett.                 B {\bf 651},      108    (2007).
\bibitem{Ripani_03}       M. Ripani et al.,                             Phys. Rev. Lett.              {\bf  91},   022002    (2003).
\bibitem{Ahrens_03}       J. Ahrens et al.,                             Phys. Lett.                 B {\bf 551},       49    (2003).
\bibitem{Ahrens_05}       J. Ahrens et al.,                             Phys. Lett.                 B {\bf 624},      173    (2005).
\bibitem{Ahrens_07}       J. Ahrens et al.,                             Eur. Phys. J.               A {\bf  34},       11    (2007).
\bibitem{Ahrens_11}       J. Ahrens et al.,                             Eur. Phys. J.               A {\bf  47},       36    (2011).
\bibitem{Krambrich_09}    D. Krambrich et al.,                          Phys. Rev. Lett.              {\bf 103},   052002    (2009).
\bibitem{Strauch_05}      S. Strauch et al.,                            Phys. Rev. Lett.              {\bf  95},   162003    (2005).
\bibitem{Nacher_01}       J.C. Nacher et al.,                           Nucl. Phys.                 A {\bf 695},      295    (2001).
\bibitem{Nacher_02}       J.C. Nacher and E. Oset,                      Nucl. Phys.                 A {\bf 697},      372    (2002).
\bibitem{Fix_05}          A. Fix and H. Ahrenh\"ovel,                   Eur. Phys. J.               A {\bf  25},      115    (2005).
\bibitem{Bianchi_94}      N. Bianchi et al.,                            Phys. Lett.                 B {\bf 325},      333    (1994).
\bibitem{Krusche_01}      B. Krusche et al.                             Phys. Rev. Lett.              {\bf  86},     4764    (2001).
\bibitem{Bernard_87}      V. Bernard, U.-G. Meissner, I. Zahed,         Phys. Rev. Lett.              {\bf  59},      966    (1987).
\bibitem{Bonutti_96}      F. Bonutti et al.,                            Phys. Rev. Lett.              {\bf  77},      603    (1996).
\bibitem{Bonutti_99}      F. Bonutti et al.,                            Phys. Rev.                  C {\bf  60},   018201    (1999).
\bibitem{Bonutti_00}      F. Bonutti et al.,                            Nucl. Phys.                 A {\bf 677},      213    (2000).
\bibitem{Camerini_04}     P. Camerini et al.,                           Nucl. Phys.                 A {\bf 735},       89    (2004).
\bibitem{Grion_05}        N. Grion et al.,                              Nucl. Phys.                 A {\bf 763},       80    (2005).
\bibitem{Starostin_00}    A. Starostin et al.,                          Phys. Rev. Lett.              {\bf  85},     5539    (2000).
\bibitem{Messchendorp_02} J.G. Messchendorp et al.,                     Phys. Rev. Lett.              {\bf  89},   222302    (2002).
\bibitem{Bloch_07}        F. Bloch et al.,                              Eur. Phys. J.               A {\bf  32},      219    (2007).
\bibitem{Herminghaus_83}  H. Herminghaus et al.,                        IEEE Trans. on Nucl. Science. {\bf  30},     3274    (1983).
\bibitem{Walcher_90}      Th. Walcher,                                  Prog. Part. Nucl. Phys.       {\bf  24},      189    (1990).
\bibitem{Anthony_91}      I. Anthony et al.,                            Nucl. Inst. and Meth.       A {\bf 301},      230    (1991).
\bibitem{Hall_96}         S.J. Hall, G.J. Miller, R. Beck, P.Jennewein, Nucl. Inst. and Meth.       A {\bf 368},      698    (1996).
\bibitem{Starostin_01}    A. Starostin et al.,                          Phys. Rev.                  C {\bf  64},   055205    (2001).
\bibitem{Novotny_91}      R. Novotny,                                   IEEE Trans. Nucl. Sci.        {\bf  38},      379    (1991).
\bibitem{Gabler_94}       A.R. Gabler et al.,                           Nucl. Inst. and Meth.       A {\bf 346},      168    (1994).
\bibitem{Watts_04}        D. Watts, in {\em Calorimetry in Particle Physics, Proceedings of the 11th International Conference, Perugia,
                          Italy 2004}, edited by C. Cecchi, P. Cenci, P. Lubrano, and M. Pepe (World Scientific, Singapore, 2005, p. 560). 
\bibitem{Audit_91}        G. Audit et al.,                              Nucl. Inst. and Meth.       A {\bf 301},      473    (1991).
\bibitem{Schumann_10}     S. Schumann et al.,                           Eur. Phys. J.               A {\bf  43},      269    (2010).
\bibitem{Olsen_59}        H. Olsen and L.C. Maximon,                    Phys. Rev.                    {\bf 114},      887    (1959).
\bibitem{Zehr_10}         F. Zehr,                                      PhD thesis, University of Basel, 2010, unpublished.
\bibitem{PDG}             K. Nakamura et al.,                           Journal of Physics          G {\bf  37},   075021    (2010).
\bibitem{Brun_86}         R. Brun et al.,                               GEANT, Cern/DD/ee/84-1,                       	     (1986). 
\bibitem{McNicoll_10}     E.F. McNicoll et al.,                         Phys. Rev.                  C {\bf  82},   035208    (2010). 
\bibitem{Pheron_12}       F. Pheron et al.,                             Phys. Lett.                 B {\bf 709},       21    (2012).
\bibitem{Kashevarov_12}   V.L. Kashevarov et al.,                       Phys. Rev.                  C {\bf  85},   064610    (2012).
\bibitem{Prakhov_09}      S. Prakhov et al.,                            Phys. Rev.                  C {\bf  79},   035204    (2009).
\bibitem{Roberts_05}      W. Roberts and T. Oed,                        Phys. Rev.                  C {\bf  71},   055201    (2005).
\bibitem{Roca_05}         L. Roca,                                      Nucl. Phys.                 A {\bf 748},      192    (2005).
\end{thebibliography}
\end{document}